\documentclass[11pt]{article}
\usepackage[T1]{fontenc}
\usepackage{graphicx} 
\usepackage{color}
\usepackage[nice]{nicefrac}
\usepackage{amsmath,amssymb}
\usepackage{comment}
\usepackage[hidelinks]{hyperref}
\usepackage{appendix}
\usepackage[normalem]{ulem}
\usepackage[nosort]{cite}

\def\hybrid{\topmargin -30pt    \oddsidemargin 0pt 
        \headheight 0pt \headsep 0pt
        \textwidth 6.25in       
        \textheight 9.5in       
        \marginparwidth .875in
        \parskip 5pt plus 1pt   \jot = 1.5ex}

\hybrid

\catcode`\@=11

\def\marginnote#1{}
\newcount\hour
\newcount\minute
\newtoks\amorpm
\hour=\time\divide\hour by60
\minute=\time{\multiply\hour by60 \global\advance\minute by-\hour}
\edef\standardtime{{\ifnum\hour<12 \global\amorpm={am}%
        \else\global\amorpm={pm}\advance\hour by-12 \fi
        \ifnum\hour=0 \hour=12 \fi
        \number\hour:\ifnum\minute<10 0\fi\number\minute\the\amorpm}}
\edef\militarytime{\number\hour:\ifnum\minute<10 0\fi\number\minute}

\def\draftlabel#1{{\@bsphack\if@filesw {\let\thepage\relax
   \xdef\@gtempa{\write\@auxout{\string
      \newlabel{#1}{{\@currentlabel}{\thepage}}}}}\@gtempa
   \if@nobreak \ifvmode\nobreak\fi\fi\fi\@esphack}
        \gdef\@eqnlabel{#1}}
\def\@eqnlabel{}
\def\@vacuum{}
\def\draftmarginnote#1{\marginpar{\raggedright\scriptsize\tt#1}}

\def\draft{\oddsidemargin -.5truein
        \def\@oddfoot{\sl preliminary draft \hfil
        \rm\thepage\hfil\sl\today\quad\militarytime}
        \let\@evenfoot\@oddfoot \overfullrule 3pt
        \let\label=\draftlabel
        \let\marginnote=\draftmarginnote
   \def\@eqnnum{(\theequation)\rlap{\kern\marginparsep\tt\@eqnlabel}%
\global\let\@eqnlabel\@vacuum}  }

\def\draft2{
        \def\@oddfoot{\sl preliminary draft \hfil
        \rm\thepage\hfil\sl\today\quad\militarytime}
        \let\@evenfoot\@oddfoot \overfullrule 3pt
        \let\marginnote=\draftmarginnote
   \def\@eqnnum{(\theequation)\rlap{\kern\marginparsep\tt\@eqnlabel}%
\global\let\@eqnlabel\@vacuum}  }

\def\preprint{\twocolumn\sloppy\flushbottom\parindent 2em
        \leftmargini 2em\leftmarginv .5em\leftmarginvi .5em
        \oddsidemargin -.5in    \evensidemargin -.5in
        \columnsep .4in \footheight 0pt
        \textwidth 10.in        \topmargin  -.4in
        \headheight 12pt \topskip .4in
        \textheight 6.9in \footskip 0pt
        \def\@oddhead{\thepage\hfil\addtocounter{page}{1}\thepage}
        \let\@evenhead\@oddhead \def\@oddfoot{} \def\@evenfoot{} }

\def\numberbysection{\@addtoreset{equation}{section}
        \def\theequation{\thesection.\arabic{equation}}}

\def\underline#1{\relax\ifmmode\@@underline#1\else
        $\@@underline{\hbox{#1}}$\relax\fi}

\def\titlepage{\@restonecolfalse\if@twocolumn\@restonecoltrue\onecolumn
     \else \newpage \fi \thispagestyle{empty}\c@page\z@
        \def\thefootnote{\fnsymbol{footnote}} }

\def\endtitlepage{\if@restonecol\twocolumn \else \newpage \fi
        \def\thefootnote{\arabic{footnote}}
        \setcounter{footnote}{0}}  

\catcode`@=12
\relax

\def\figcap{\section*{Figure Captions\markboth
        {FIGURECAPTIONS}{FIGURECAPTIONS}}\list
        {Figure \arabic{enumi}:\hfill}{\settowidth\labelwidth{Figure
999:}
        \leftmargin\labelwidth
        \advance\leftmargin\labelsep\usecounter{enumi}}}
 \relax
\def\tablecap{\section*{Table Captions\markboth
        {TABLECAPTIONS}{TABLECAPTIONS}}\list
        {Table \arabic{enumi}:\hfill}{\settowidth\labelwidth{Table
999:}
        \leftmargin\labelwidth
        \advance\leftmargin\labelsep\usecounter{enumi}}}
 \relax
\def\reflist{\section*{References\markboth
        {REFLIST}{REFLIST}}\list
        {[\arabic{enumi}]\hfill}{\settowidth\labelwidth{[999]}
        \leftmargin\labelwidth
        \advance\leftmargin\labelsep\usecounter{enumi}}}
 \relax
\makeatletter
\newcounter{pubctr}
\def\publist{\@ifnextchar[{\@publist}{\@@publist}}
\def\@publist[#1]{\list
        {[\arabic{pubctr}]\hfill}{\settowidth\labelwidth{[999]}
        \leftmargin\labelwidth
        \advance\leftmargin\labelsep
        \@nmbrlisttrue\def\@listctr{pubctr}
        \setcounter{pubctr}{#1}\addtocounter{pubctr}{-1}}}
\def\@@publist{\list
        {[\arabic{pubctr}]\hfill}{\settowidth\labelwidth{[999]}
        \leftmargin\labelwidth
        \advance\leftmargin\labelsep
        \@nmbrlisttrue\def\@listctr{pubctr}}}
 \relax
\makeatother

\newcommand{\beq}{\begin{equation}}
\newcommand{\eeq}{\end{equation}}

\newcommand{\m}{\mu}
\newcommand{\n}{\nu}
\newcommand{\prof}[1]{\begin{proof}[\textbf{Proof of }\eqref{#1}]}

\begin{document}

\renewcommand{\theequation}{\thesection.\arabic{equation}}
\csname @addtoreset\endcsname{equation}{section}

\begin{titlepage}
\begin{center}


\phantom{xxx}
\vskip 0.5in

{\huge \bf On the geometric origin of the energy--momentum tensor improvement terms}

\vskip 0.5in

{\bf Damianos Iosifidis}${}^{1a}$,\phantom{x} {\bf Manthos Karydas}${}^{2b}$\\
{\bf Anastasios Petkou}${}^{3c}$\phantom{x}and\phantom{x}{\bf Konstantinos Siampos}${}^{3d}$ \vskip 0.1in

${}^1$Laboratory of Theoretical Physics, Institute of Physics,\\
University of Tartu, W. Ostwaldi 1, 50411 Tartu, Estonia\\


\vskip 0.11in

${}^2$Physics Division, Lawrence Berkeley National Laboratory, Berkeley, CA, USA


${}^3$Laboratory of Theoretical Physics, School of Physics,\\
Aristotle University of Thessaloniki, 54124 Thessaloniki, Greece

\vskip .2in

 {\footnotesize
$^a$damianos.iosifidis@ut.ee, $^b$mkarydas@berkeley.edu, $^c$petkou@auth.gr and $^d$ksiampos@auth.gr}

\vskip .2in


\end{center}

\vskip .4in

\centerline{\bf Abstract}
\vskip .1in
\noindent
In a flat background, the canonical energy momentum tensor of Lorentz and conformally invariant matter field theories 
can be improved to a symmetric and traceless tensor that gives the same conserved charges. We argue that the geometric origin of this improvement process is unveiled when the matter theory is coupled to Metric-Affine Gravity. In particular, we show that the Belinfante--Rosenfeld improvement terms correspond to the matter theory's hypermomentum. The improvement terms in conformally invariant matter theories are also related to the hypermomentum however a general proof would require an extended investigation. We demonstrate our results through various examples, such as the free massless scalar, the Maxwell field, Abelian $p$-forms, the Dirac field and a non-unitary massless scalar field. Possible applications of our method for theories that break Lorentz or special conformal invariance are briefly discussed.

\end{titlepage}

\def\baselinestretch{1}
\baselineskip 20 pt

\tableofcontents


\section{Introduction}

The energy momentum (e.m.) tensor is a fundamental quantity in any field theory as it is associated with its symmetries and the corresponding conserved charges. Nevertheless, even on a fixed flat background the e.m.  of a matter action is not uniquely defined although its different forms give rise to the same conserved charges. For example, given a translation invariant action one constructs by the standard Noether's procedure the so-called {\it canonical} e.m. tensor $t^{\mu\nu}$ that yields the conserved energy and spatial momentum of the theory. However, if the action is also Lorentz invariant  the canonical e.m. tensor satisfies a certain property which enables one to define an "improved" e.m. tensor, the so-called Belinfante--Rosenfeld (BR) e.m. tensor $T_\text{BR}^{\mu\nu}$, which is symmetric in $\mu,\nu$~\cite{Belinfante,Rosenfeld}. Both $t^{\mu\nu}$ and $T_\text{BR}^{\mu\nu}$ give rise to the same conserved charges e.g.~\cite{Blaschke:2016ohs}, i.e. the relativistic angular momenta, however the symmetry property of the latter generically makes it more preferable {  as the latter is explicitly symmetric under Lorentz transformations}. 

Similarly, for Lorentz invariant theories that are also invariant under scale transformations the Noether's procedure yields a conserved dilatation current which is given by a combination of the canonical e.m. tensor and a "virial" current~\cite{Wess:1960}. If the  theory is also invariant under special conformal transformations, the canonical e.m. tensor satisfies a certain property which enables one to "improve" it and define a conserved, symmetric and traceless e.m. tensor $\Theta^{\mu\nu}$~\cite{Gross:1970tb}. Both $t^{\mu\nu}$ and $\Theta^{\mu\nu}$ give rise to the same conserved charges however the tracelesness of the latter generically makes it more preferable as the latter is explicitly symmetric under scale and special conformal transformations. 

Perhaps the most important property of the improved e.m. tensors $T_\text{BR}^{\mu\nu}$ and $\Theta^{\mu\nu}$ is that these quantities are somehow "chosen" by Einstein gravity when the latter couples to matter.  Indeed, the variation wrt to the metric of any diffeomorphic invariant extension of a Lorentz invariant matter action defines a symmetric and covariantly conserved e.m. tensor, the Hilbert or {\it metrical}  e.m. tensor $T^{\m\n}$. This tensor generically coincides with the improved BR e.m. tensor $T_\text{BR}^{\mu\nu}$ in the usual flat limit. However, if the flat action is conformally invariant, one might want to couple it to Einstein gravity in a Weyl invariant manner~\cite{Callar.Jackiw}. It turns out that the metrical e.m. tensor of a Weyl invariant action is also traceless and it coincides with the improved conformal one $\Theta^{\m\n}$ in the usual flat limit. 

Hence, coupling matter to gravity appears to circumvent the improvement process for the canonical e.m. tensor of matter theories in a flat backgrounds. This is expected since coupling a flat matter theory to gravity is equivalent to making it diffeomorphic (and Weyl) invariant and hence to the gauging of all of its global flat spacetime symmetries. Nevertheless, it is interesting to ask whether there is a gravitational/geometrical analog of the improvement process that leads from the canonical flat e.m. tensor $t^{\m\n}$ to $T_\text{BR}^{\m\n}$ and $\Theta^{\m\n}$. Motivated by this question, we study in this work the coupling of flat matter actions to Metric-Affine Gravity (MAG)~\cite{Hehl:1976my}, see also~\cite{Hehl:1994ue} for a review. Such a wider geometrization gives rise not only to a metrical e.m. tensor $T^{\m\n}$ of the curved action, but also to the {\it hypermomentum} $\Delta_\lambda{}^{\m\n}$ defined as the variation of the action wrt to the affine connection)~\cite{Hehl:1976kv,Hehl:1977gn}. In this case diffeomorphism invariance leads to a more complicated conservation equation that involves both $T^{\m\n}$ and $\Delta_\lambda{}^{\m\n}$. Furthermore, MAG extensions of Weyl invariance lead to trace identities that also involve both $T^{\m\n}$ and $\Delta_\lambda{}^{\m\n}$ \cite{Iosifidis:2018zwo}\footnote{Trace anomaly in MAG has been recently studied in \cite{Bahamonde:2024kyi}.}. Our observation is that the usual flat limits of the above conservation and trace identities differ from the corresponding ones of Einstein gravity due to the presence of the hypermomentum.  This means that the flat limit of the MAG metrical e.m. tensor does not generically coincide neither with the BR e.m. tensor $T_\text{BR}^{\m\n}$ nor with the improved 
conformal one $\Theta^{\m\n}$~\cite{Polchinski:1987dy}. This in turn implies that the hypermomentum is intimately related to the improvement terms of the canonical flat space e.m. tensor $t^{\m\n}$. The explicit relationship of the hypermomentum $\Delta_\lambda{}^{\m\n}$ to the improvement terms is theory dependent as we demonstrate in number of examples that include scalars, fermions and vector fields. 

The paper is organized as follows. In Section \ref{Sec2} we briefly review the improvement procedure for the canonical e.m. tensor of matter theories in flat space. In Section \ref{setup} we study matter fields in MAG backgrounds and write down the conservation and trace conditions that follow from diffeomorphism and MAG-Weyl invariance. In Section \ref{rest.Lorentz} we present examples of Lorentz invariant flat matter theories where the improvement process of their canonical e.m. tensor is captured by the hypermomentum. These include the Maxwell field, Dirac fermions and higher-form abelian matter fields. In Section \ref{rest.conformal} we present examples of conformally invariant flat matter theories whose improvement process for their canonical e.m. tensor is also captured by the hypermomentum. These include canonical scalars as well as higher-derivative massless scalars. We summarize our results and discuss their possible generalization in Section \ref{Conclusion.Outlook}. In Appendix \ref{Appendix.MAG}, we summarize the metric-affine framework. Then, in Appendices \ref{Appendix.Canonical} and \ref{Appendix.proof} we revisit the canonical stress-tensor as a variation with respect to the vielbein and derive its expression with the metric-affine theories. In Appendix \ref{Appendix.Examples}, we revisit the improvement terms in the various examples considered in the main text. Finally, in Appendix~\ref{Sec:Improved} we revisit improvements of the conserved current and we apply it in the various cases studied in the main text. 

\section{The improved energy momentum tensors of matter fields in a flat background}
\label{Sec2}

We briefly review here the well-known improvement process for matter fields in a flat background. This discussion can be found in many places in the literature and we follow \cite{Gross:1970tb}.\footnote{Generalizations of the Noether method have been considered in~\cite{Osborn.CFTs,Brauner:2019lcb,Freese:2021jqs,Kourkoulou:2022ajr,Gieres:2022cpn,Kim:2024ewt}.} We consider matter fields $\phi^A(x)$, where $A$ collectively denotes spacetime or internal symmetry indices, described by the second order action
\beq
\label{Sflat}
S[\phi^A]=\int d^dx\,{\cal L}(\phi^A,\partial_\mu\phi^A)\,.
\eeq
The metric is Minkowski with a mostly-plus signature. A straightforward way to study the global spacetime symmetries of the action \eqref{Sflat} is to consider general {\it active} coordinate transformations of the form $x^\m\mapsto x^\m-\xi^\m(x)$ that change the action by a total derivative as\footnote{This follows from the fact that the Lagrangian  ${\cal L}(x)$ is just a scalar field.}
\beq
\label{dSflat1}
S[\phi^A]\mapsto S[\phi^A]+\delta^{(1)}_\xi S[\phi^A]=S[\phi^A]+\int d^dx\,\partial_\m\left(\xi^\m{\cal L}(\phi^A,\partial\phi^A)\right)\,.
\eeq
Next, the fields transform under the active transformation as
\beq
\label{dphi}
\phi^A\mapsto \phi^A+\delta_\xi\phi^A=\phi^A+\xi^\m\partial_\m\phi^A+\Delta_\xi\phi^A\,.
\eeq
We split $\delta_\xi\phi^A$ into a first term that it is always there and a second term, $\Delta_\xi\phi^A$, that depends on the tensorial properties of $\phi^A$ as well as on the form of the transformation. We further require that $\Delta_\xi\phi^A$ is at most linear in whatever small parameter is inside $\xi^\mu$. Under \eqref{dphi} the action \eqref{Sflat} transforms as
\begin{eqnarray}
\label{dSflat2}
&S[\phi^A]\mapsto &S[\phi^A]+\delta^{(2)}_\xi S[\phi^A]=S[\phi^A]+
\int d^dx\left(\frac{\partial{\cal L}}{\partial\phi^A}\delta_\xi\phi^A+\frac{\partial{\cal L}}{\partial(\partial_\m\phi^A)}\partial_\m(\delta_\xi\phi^A)\right)\,.
\end{eqnarray}
Requiring that $[\delta^{(1)}_\xi-\delta^{(2)}_\xi]S[\phi^A]=0$ gives
\beq
\label{d12}
\int d^dx\left[\partial_\m\left(\xi^\n t^\m_{\,\,\n}+\pi^{\m,A}\Delta_\xi\phi^A\right)+\delta_\xi\phi^A\times \text{e.o.m.}\right]=0\,.
\eeq
where the canonical e.m. tensor $t^\m{}_{\n}$ is defined as
\beq
\label{tmnpi}
t^\m{}_{\n}=\pi^{\m,A}\partial_\n\phi^A-\delta^\m_{\,\,\n}{\cal L}\,,\,\,\,\,
\pi^{\m,A}=\frac{\partial{\cal L}}{\partial(\partial_\m\phi^A)}\,,
\eeq
and the equations of motion (e.o.m.) are as usual
\beq
\label{eom}
\text{e.o.m.}=\frac{\partial{\cal L}}{\partial\phi^A}-\partial_\m\left(\frac{\partial{\cal L}}{\partial(\partial_\m\phi^A)}\right)\,.
\eeq
Hence, on-shell we obtain the well-known (i.e. see \cite{Gross:1970tb})  relation the yields the conserved currents for a spacetime symmetry in flat space
\beq
\label{cEq}
\partial_\m\left(\xi^\n t^\m{}_{\n}+\pi^{\m,A}\Delta_\xi\phi^A\right)=0\,.
\eeq
From (\ref{cEq}) we can read the conditions satisfied by the canonical e.m. tensor $t^{\m}_{\,\,\n}$ which lead to the improvement process. 
\begin{itemize}
\item
Translations correspond to $\xi^\m=a^\m$ with $a^\m$ constant. This gives
\beq
\label{transl}
\Delta_a\phi^A=0\,\,\Rightarrow\,\,\partial_\m t^\m_{\,\,\n}=0\,.
\eeq
\item
Lorentz rotations correspond to 
\beq
\label{Lorentz.transf}
\xi^\m=\omega^\m{}_{\n}x^\n\,,\quad \omega^{\m\n}=-\omega^{\n\m}=\text{constants}\,.
\eeq 
This gives
\beq
\label{Lorentz1}
\Delta_{\omega}\phi^A=\omega_{\m\n}\Sigma^{\m\n,A}\,,
\eeq
where $\Sigma^{\m\n,A}=-\Sigma^{\n\m,A}$ is a function of the field $\phi^A$ that depends on their Lorentz properties. Substituting \eqref{Lorentz1} into \eqref{cEq} we obtain\footnote{Where we introduce our conventions for symmetrizing and antisymmetrizing 
$A_{(\mu\nu)}=
\frac12(A_{\mu\nu}+A_{\nu\mu})$
and
$A_{[\mu\nu]}=
\frac12(A_{\mu\nu}-A_{\nu\mu})$.}
\beq
\label{Lorentz2}
t^{\m\n}-t^{\n\m}=2\partial_\rho\Psi^{\rho[\m\n]}\,,
\eeq
where the quantity $\Psi^{\rho[\m\n]}=\pi^{\rho,A}\Sigma^{[\m\n],A}$ 
is referred to as the {\it spin tensor}. The condition \eqref{Lorentz2} satisfied by the antisymmetric part of the canonical e.m. tensor is the imprint of Lorentz invariance of the action \eqref{Sflat}. It immediately leads to the well-known BR improvement procedure as~\cite{Belinfante,Rosenfeld}
\beq
\label{BR1}
t^{\m\n}=T^{\m\n}_\text{BR}+\partial_\rho\left(\Psi^{\m[\n\rho]}+\Psi^{\n[\m\rho]}+\Psi^{\rho[\m\n]}\right)\,,
\eeq
where the BR e.m. tensor $T^{\m\n}_\text{BR}$ is symmetric and on-shell conserved
\beq
\label{BR2}
\partial_\m T^{\m\n}_\text{BR}=0\,,\,\,\,\,T^{\m\n}_\text{BR}=T^{\n\m}_\text{BR}\,.
\eeq
\item
Dilatations correspond to $\xi^\m=\lambda x^\m$ with $\lambda$ constant and hence 
\beq
\label{dil1}
\Delta_\lambda\phi^A=\lambda \Sigma^A\,,
\eeq
for a function $\Sigma^A$ of the fields. Then from (\ref{cEq}) we obtain
\beq
\label{dil2}
t^{\m}{}_{\m}+\partial_\m N^\m=0\,,
\eeq
with $N^\m=\pi^{\m,A}\Sigma^A$ referred to as the {\it virial current}. If the theory is Lorentz invariant \eqref{dil2} implies a condition for the trace of the Bellinfante--Rosenfeld e.m. tensor as
\beq
\label{dil3}
T^{\,\,\m}_{\text{BR}\,\m}+\partial_\m\left(N^\m+2\Psi_\n^{\,\,[\n\m]}\right)=0\,.
\eeq
\item
Special conformal transformations correspond to $\xi^\m=x^2b^\m-2b^\n x_\n x^\m$ with $b^\m$ constants and hence
\beq
\label{sconf1}
\Delta_b\phi^A=b_\m\Sigma^{\m,A}\,,
\eeq
for a function $\Sigma^{\m,A}$ of the fields. Substituting (\ref{sconf1}) into (\ref{cEq}) we obtain
\beq
2x_\m t^{[\m\n]}-2x^\n t^{\m}_{\,\m}+\partial_\m N^{\m\n}=0\,,
\eeq
with $N^{\m\n}=\pi^{\m,A}\Sigma^{\n,A}$. 
Using (\ref{Lorentz2}) and (\ref{dil2}) we then find
\beq
\label{sconf2}
N^\m+2\Psi_\n{}^{\,[\n\m]}=\partial_\n\Lambda^{\m\n}\,,
\eeq
with $2\Lambda^{\m\n}=N^{\m\n}+2x^\n N^\m+2x_\rho\Psi^{\m[\rho\n]}$. Finally, assuming Lorentz invariance we can use \eqref{dil3} to obtain the condition satisfied by the BR e.m. tensor of a conformally invariant theory as~\cite{Polchinski:1987dy,Conformal1}
\beq
\label{sconf4}
T^{\,\,\m}_{\text{BR}\,\m}+\partial_\m\partial_\n\Lambda^{\m\n}=0\,.
\eeq
Using $\Lambda^{\m\n}$ it is then possible to construct a conserved, symmetric, and traceless e.m. tensor $\Theta^{\m\n}$ as in~\cite{Polchinski:1987dy}
\begin{equation}
\begin{split}
\label{sconf5}
\Theta^{\m\n}&=T^{\m\n}_\text{BR}-\frac{1}{d-2}\left(\partial^\m\partial_\rho\Lambda^{\rho\n}+\partial^\n\partial_\rho\Lambda^{\rho\m}-\partial^2\Lambda^{\m\n}-\eta^{\m\n}\partial_\rho\partial_\sigma\Lambda^{\rho\sigma}\right)\\
&-\frac{1}{(d-1)(d-2)}\left(\eta^{\m\n}\partial^2-\partial^\m\partial^\n\right)\Lambda^\rho{}_{\rho}\,,
\end{split}
\end{equation}
where the additional term in \eqref{sconf5} is identically conserved and 
\beq
\label{sconf6}
\partial_\m\Theta^{\m\n}=\Theta^\m_{\,\m}=0\,,\,\,\,\Theta^{\m\n}=\Theta^{\n\m}\,.
\eeq
The dilation and special conformal conserved currents are now nicely written as~\cite{Conformal1}
\beq
\label{currents.scale.special}
j^\mu_D=x_\nu \Theta^{\mu\nu}\Rightarrow \partial_\mu j^\mu_D=0\,,\quad K^{\mu\nu}=x^2\Theta^{\mu\nu}-2x^\nu x_\kappa \Theta^{\mu\kappa}\Rightarrow \partial_\mu K^{\mu\nu}=0
 \,.
\eeq
\end{itemize}
To recap, calculating directly from the Lagrangian the canonical e.m. tensor $t^{\m\n}$ allows us to deduce whether the matter theory is Lorentz, dilatation and conformal invariant by checking the conditions \eqref{Lorentz2}, \eqref{dil2} and \eqref{sconf2}. Then, we can construct accordingly its corresponding improved versions.

\section{Matter fields in curved backgrounds}
\label{setup}
\subsection{Matter fields coupled to Einstein gravity}

Coupling the matter theory \eqref{Sflat} to Einstein gravity is usually done by promoting the partial derivatives to covariant ones $\partial_\m\mapsto\nabla_\m$. Although this procedure is generally ambiguous due to the fact that the flat action \eqref{Sflat} can be modified by total derivatives terms, it yields an unambiguous result regarding the e.m. tensor. Namely, suppose that we extend \eqref{Sflat} to curved space as
\beq
\label{Scurved}
 S[\phi^A]\mapsto S[\phi^A;g]=\int d^dx\sqrt{-g}\,{\cal L}(\phi^A,\nabla_\m\phi^A;g)\,.
 \eeq
Under general diffeomorphisms $x^\m\rightarrow x^\m-\xi^\m$ this transforms as
 \begin{eqnarray}
 \label{dScurved}
 \delta_\xi S[\phi^A;g]&=&-\frac{1}{2}\int d^d x\sqrt{-g}\,T^{\m\n}{\cal L}_\xi g_{\m\n}\,,\nonumber \\
 &=&\int d^dx\sqrt{-g}\,\xi_\n\nabla_\m T^{\m\n}-\int d^d x\partial_\m(\sqrt{-g}\xi_\n T^{\m\n})\,.
 \end{eqnarray}
 As we are interested in the on-shell variation of the matter action we have not indicated the additional terms that are proportional to the e.o.m. of the matter fields.
As usual the Lie derivative of the metric is ${\cal L}_\xi g_{\m\n}=\nabla_\m\xi_\n+\nabla_\n\xi_\n$ and 
the symmetric Hilbert metrical e.m. tensor $T^{\m\n}$ is defined as
\beq
\label{grav.stress}
T_{\mu \nu} := - \frac{2}{\sqrt{-g}} \frac{\delta S}{\delta g^{\mu \nu}}\,,\,\,\,\,\quad g:=\det g_{\mu\nu}\,.
\eeq
If we require that (\ref{Scurved}) is invariant under diffeomorphisms, this means that its variation vanishes up to boundary terms and hence we obtain the on-shell conservation equation 
\beq
\label{conT}
\nabla_\m T^{\m\n}=0\,.
\eeq
Restricting then back to flat space we see that $T^{\m\n}$  correspond to a symmetric and conserved e.m. tensor and might only differ from the improved BR e.m. tensor $T_\text{BR}^{\m\n}$ by terms that do not alter the Poincare charges. 

If the flat matter theory on top of being Lorentz invariant is conformally invariant as well, then we may wish to couple it to Einstein gravity in a Weyl invariant manner \cite{Callar.Jackiw}. Namely we might want to find a diffeomorphic invariant action which is also invariant under simultaneous Weyl rescaling of the metric $\delta_\Omega g_{\m\n}=2\Omega g_{\m\n}$ and of the matter fields\footnote{The Weyl rescaling of matter fields does not depend on their spin.}  $\delta_\Omega\phi^A=\Delta_\phi \Omega\phi^A$ with $\Delta_\phi$ the scaling dimension. The resulting curved action $S_w[\phi^A;g_{\m\n}]$ is not the same as \eqref{Scurved}, nevertheless it gives rise to a metrical e.m. tensor for the $S_w$ action which is symmetric and covariantly conserved. However, Weyl invariance of the action gives 
\beq
\label{dSWeyl}
\delta_\Omega S_w[\phi^A;g]=-\frac{1}{2}\int d^dx\sqrt{-g}T_{(w)}^{\m\n}\delta_\Omega g_{\m\n}=-\frac{1}{2}\int d^dx\sqrt{-g}\,T_{(w)}^{\m\n}2\Omega\, g_{\m\n}=0\,.
\eeq
This yields $g_{\m\n}T_{(w)}^{\m\n}=0$ for the metrical e.m. tensor of the Weyl invariant action. Restricting finally to flat space we see that $T_{(w)}^{\m\n}$ correspond to a conserved, symmetric and traceless e.m. tensor of a conformally invariant matter theory, hence it either coincides with the conformally improved one $\Theta^{\m\n}$, or it differs from the latter by terms that do not alter the conformal charges. This Weyl-covariantization procedure is the standard method of obtaining the tensor $\Theta^{\m\n}$ in the case of complicated matter theories i.e. \cite{Osborn.CFTs}.

\subsection{Matter fields in metric-affine  backgrounds}

It is not too surprising that Einstein and Weyl gravity couple to matter through the improved e.m. tensors $T_\text{BR}^{\m\n}$ and $\Theta^{\m\n}$ respectively. After all general covariance and Weyl covariance include  Poincar\'e and conformal symmetry. Nevertheless, this leaves the flat space improvement procedure without a clear geometric interpretation. Our aim here is to provide a geometric framework for the e.m. improvement process in flat space, which we will argue that is unveiled  through the coupling of the matter theory to metric-affine gravity (MAG).\footnote{For a reminder of metric-affine gravity and our notations see Appendix~\ref{Appendix.MAG}.}

\subsection{Conservation and trace identities}

The matter action \eqref{Sflat} coupled to MAG in a holonomic (i.e. coordinate based) takes the form
\beq
\label{SMAG}
S[\phi^A]\mapsto S[\phi^A; g,\Gamma]=\int d^{d}x \sqrt{-g}\,\mathcal{L}(\phi^A;g,\Gamma)\,.
\eeq
The fields $\phi^A$(x) couple both to the metric $g_{\m\n}$ and to the independent affine connection $\Gamma^{\lambda}{}_{\mu\nu}$. Now, apart from the metrical e.m. tensor defined in \eqref{grav.stress} the action \eqref{SMAG} gives rise to an additional quantity, the hypermomentum  \cite{Hehl:1976kv,Hehl:1977gn}
defined as the variation of the matter part of the action with respect to the affine connection
\beq
\label{hyper.def}
{\Delta_\lambda}^{\mu \nu} := - \frac{2}{\sqrt{-g}} \frac{\delta S}{\delta {\Gamma^\lambda}_{\mu \nu}}\,,\quad
\Delta^{\lambda\mu\nu}:=g^{\lambda\rho}{\Delta_\rho}^{\mu \nu}\,.
\eeq

Physically, the hypermomentum  \eqref{hyper.def} encodes the micro-properties of matter such as spin, dilation and shear. It can be split into  three pieces as (see \cite{Hehl:1994ue}) 
\beq
\Delta^{\mu\nu\rho}=\Sigma^{\mu\nu\rho}+\frac{g^{\mu\nu}}{d}\Delta^{\rho}+\hat{\Delta}^{\mu\nu\rho} \,,
\label{hypsplit}
\eeq
with
\beq
\label{hyper.Spin}
\Sigma^{\mu\nu\rho}:=\Delta^{[\mu\nu]\rho} \qquad \text{Spin}
\eeq
\beq
\label{hyper.Dilation}
\Delta^{\nu}:=\Delta^{\alpha\mu\nu}g_{\alpha\mu} \qquad \text{Dilation}
\eeq
\beq
\hat\Delta^{\mu\nu\rho}:=\Delta^{(\mu\nu)\rho}-\frac{1}{d}g^{\mu\nu}\Delta^{\rho} \qquad \text{Shear}
\eeq

As in \eqref{dScurved},  the on-shell variation of the action \eqref{SMAG} under  the diffeomorphisms  $x^\mu=x'^\mu+\xi^\mu$ is given by 
\beq
\label{variation.action}
\delta_\xi S=-\frac12\int d^{d}x \sqrt{-g}\left(T_{\mu\nu}\,{\cal L}_\xi g^{\mu\nu}+
\Delta_\lambda{}^{\mu\nu}\,{\cal L}_\xi \Gamma^\lambda{}_{\mu\nu}\right)\,.
\eeq
The Lie derivatives of the metric and the affine connection are
\begin{eqnarray}
{\cal L}_\xi g_{\mu\nu}&=&
g_{\rho\nu}\partial_\mu \xi^\rho+
g_{\mu\rho}\partial_\nu \xi^\rho+\xi^\rho \partial_\rho g_{\mu\nu}\\
&=&g_{\nu\rho}\nabla_\mu \xi^\rho+
g_{\mu\rho}\nabla_\nu \xi^\rho-2\xi^\rho(S_{\rho\mu\nu}
+S_{\rho\nu\mu})-\xi^\rho Q_{\rho\mu\nu}\,,\nonumber\\
\label{ksjksks}
{\cal L}_\xi\Gamma^\mu{}_{\kappa\lambda}&=&
-\partial_\rho \xi^\mu\Gamma^\rho{}_{\kappa\lambda}+
\partial_\kappa\xi^\rho \Gamma^\mu{}_{\rho\lambda}
+\partial_\lambda\xi^\rho \Gamma^\mu{}_{\kappa\rho}
+\xi^\rho\partial_\rho\Gamma^\mu{}_{\kappa\lambda}+\partial_\kappa\partial_\lambda\xi^\mu\nonumber\\
&=&\nabla_\kappa\left(2\xi^\rho S_{\rho\lambda}{}^\mu+\nabla_\lambda\xi^\mu\right)-R^\mu{}_{\lambda\kappa\rho}\,\xi^\rho\,.
\end{eqnarray}
The last equality conforms that well-known fact that the Lie derivative of the affine-connection is a tensor.
Plugging the above into \eqref{variation.action}, we find~\cite{Hehl:1989ij,Hehl:1994ue} 
\begin{equation}
\label{dksfslkkq}
\delta_\xi S=\int d^dx\, \xi^\alpha \Xi_\alpha+\int d^{d}x \partial_{\mu}(\sqrt{-g}j^{\mu})\,,
\end{equation}
with 
\begin{equation}
\label{ccc0}
\Xi_\alpha=\sqrt{-g}(2 \tilde{\nabla}_{\mu}T^{\mu}_{\;\;\alpha}-\Delta^{\lambda\mu\nu}R_{\lambda\mu\nu\alpha})+\hat{\nabla}_{\mu}\hat{\nabla}_{\nu}(\sqrt{-g}\Delta_{\alpha}^{\;\;\mu\nu})-2S_{\mu\alpha}^{\;\;\;\;\lambda}\hat{\nabla}_{\nu}(\sqrt{-g}\Delta_{\lambda}^{\;\;\;\mu\nu})\,,
\end{equation}
where we introduce $\hat{\nabla}_{\lambda}=\nabla_{\lambda}-2 S_{\lambda}$ \footnote{The $\nabla$ covariant derivative acts on the scalar density as 
$\nabla_\lambda\sqrt{-g}=-\frac12 Q_\lambda \sqrt{-g}$ and trivially follows
$\hat\nabla_\lambda\sqrt{-g}=-\left(\frac12 Q_\lambda+2S_\lambda\right) \sqrt{-g}$, see also equation~\eqref{nabla.density}.}  and 
\beq
\label{conserved.j}
j^{\mu}=-\xi_{\nu}\left(T^{\mu\nu}+\frac{g^{\nu\rho}}{2 \sqrt{-g}}\hat{\nabla}_\lambda(\sqrt{-g}\Delta_{\rho}{}^{\mu\lambda})\right)
+\frac{1}{2}\Delta_{\lambda}{}^{\nu\mu}\nabla_{\nu}\xi^{\lambda}+S_{\nu\rho}{}{}^{\lambda}\xi^{\rho}\Delta_{\lambda}{}^{\nu\mu}\,.
\eeq
Requiring that  \eqref{dksfslkkq} vanishes up to boundary terms gives the {\it generalized conservation} of metric-affine gravity \cite{Hehl:1989ij,Hehl:1994ue} 
\beq
\label{ksi}
\Xi^\mu=0\,.
\eeq
In the MAG framework one defines generalised Killing vectors $\xi^\m$ that satisfy~\cite{Hecht:1992xn}\footnote{Equivalently the generalized Killing vectors can be  defined as ${\cal L}_\xi g_{\mu\nu}=
{\cal L}_\xi S_{\mu\nu\rho}=
{\cal L}_\xi Q_{\mu\nu\rho}=0$.}
\beq
{\cal L}_\xi g_{\mu\nu}=0\,,\quad 
{\cal L}_\xi \Gamma^\lambda{}_{\mu\nu}=0\,.
\eeq
In this case along with \eqref{ksi} we obtain the conservation law~\cite{Hecht:1992xn}
\beq
\label{conserved.j1}
\partial_{\mu}(\sqrt{-g}j^{\mu})=0\,,
\eeq 
where the presence of the hypermomentum in \eqref{conserved.j} generalizes the usual definition of conserved currents in a matter theory coupled to Einstein gravity,  i.e. $\partial_{\mu}(\sqrt{-g}\xi_{\nu}T^{\mu\nu})=0$.

The generalization of Weyl invariance~\cite{Folland,Hall,Ciambelli:2019bzz} in the context of MAG in a MAG framework is not unique as one has the possibility to consider different transformations of the affine connection in conjunction with the Weyl rescaling of the metric~\cite{Iosifidis:2018zwo}. Here we will consider the transformations 
\beq
\label{connection.Weyl}
g_{\mu\nu}\mapsto e^{-2\Omega}g_{\mu\nu}\,,\quad
\Gamma^\rho{}_{\mu\nu}\mapsto
\Gamma^\rho{}_{\mu\nu}-\partial_\nu\Omega\delta_\mu^\rho\,,
\eeq
for some arbitrary $\Omega=\Omega(x)$ which were first considered in~\cite{Nieh:1982nb,Obukhov:1982zn,Dereli:1982xb}. These transformations correspond to a {\it local frame rescaling} of the form ~\cite{Iosifidis:2018zwo} 
\beq
\label{lfr}
e_\mu{}^a\mapsto e^{-\Omega}\,e_\mu{}^a
\eeq
and the transformation of the generic affine connection in \eqref{connection.Weyl}, with the spin connection kept fixed, can be read from \eqref{spin.constraint} below.   Under frame rescaling the action  \eqref{SMAG}  transforms on-shell as
\beq
\begin{split}
\label{dSM}
\delta_\Omega S&=-\frac12\int d^{d}x \sqrt{-g}\left(T_{\mu\nu}\,\delta_\Omega g^{\mu\nu}+
\Delta_\lambda{}^{\mu\nu}\,\delta_\Omega \Gamma^\lambda{}_{\mu\nu}\right)\\
&=-\int d^d x\sqrt{-g}\,\Omega \left(T^{\mu}{}_{\mu}+\frac{1}{2 \sqrt{-g}}\hat{\nabla}_\lambda(\sqrt{-g}\Delta_{\mu}{}^{\mu\lambda})\right)+\frac12\int d^dx\partial_\nu(\sqrt{-g}\Delta_\lambda{}^{\lambda\nu}\Omega)\,.
\end{split}
\eeq
Demanding that (\ref{dSM}) vanishes up to boundary terms we obtain 
\beq
\label{MAGtrace}
T^{\mu}{}_{\mu}+\frac{1}{2 \sqrt{-g}}\hat{\nabla}_\lambda(\sqrt{-g}\Delta_{\mu}{}^{\mu\lambda})=0\,,
\eeq
that generalizes the trace condition of the metrical e.m. tensor in the case of Weyl gravity.

The simple flat limits of the conservation \eqref{ksi} and trace \eqref{MAGtrace} MAG equations do not coincide with the ones coming from coupling matter to Einstein and Weyl gravity due to the presence of the hypermomentum. For example, the only property of the flat limit of the metrical MAG e.m. tensor $T^{\m\n}$ is that it is symmetric - it is neither conserved nor traceless in general. It is only conserved when the hypermomentum is antisymmetric in the last pair of indices. Hence, it does not generically give the improved BR $T_\text{BR}^{\m\n}$ or the improved conformal  $\Theta^{\m\n}$ e.m. tensors. Nevertheless, there is an important information that we have gained from the above endeavour. We can formulate the  above matter system using the vielbein formalism and consider as independent variables, the  tangent space metric $g_{ab}$,\footnote{This can be chosen to be the Minkowski metric by an appropriate choice of the vielbeins (i.e. for orthonormal). We adopt the mostly plus convention for the latter, that is, $\eta_{ab}=\text{diag}(-1,1,\dots,1)$.} the  coframe 1-form $\vartheta^{a}=e_{\mu}{}^{a}dx^{\mu}$ and the connection (or spin connection) 1-form $\omega_{ab}=\omega_{ab\mu}dx^{\mu}$. Then the matter action functional now reads 
\beq
S[\phi^A]\mapsto S_{e}[\phi^A;e,\omega,g]=\int d^{d}x \sqrt{-g}\,\mathcal{L}_e(\phi^A;e,\omega,g)\,,
\eeq
and the additional matter source is  the {\it MAG canonical} e.m. tensor given by
\beq\label{cemt}
{\tau^a}_\mu := -\frac{1}{\sqrt{-g}} \frac{\delta S_e}{\delta {e^\mu{}_a}}\,,
\eeq
which is the fundamental e.m. tensor in MAG.
We do not consider the variation with respect to the spin-connection as it coincides with that of the connection, namely   \eqref{hyper.def}, due to the identity\footnote{Sometimes  mistakenly referred to as {\it the vielbein postulate}.}
\beq
\label{spin.constraint}
\nabla_\mu e_\nu{}^a:=\partial_\mu e_\nu{}^a+\omega^a{}_{b\mu}e_\nu{}^b
-\Gamma^\rho{}_{\nu\mu} e_\rho{}^a=0\,,
\eeq
relating the holonomic affine connection to the (anholonomic)  connection and the vielbeins.\footnote{The variation with respect to the tangent space metric gives an important relation that is a combination of the variations wrt to vielbein and spin connection \cite{Hehl:1994ue} (see below).}  The crucial observation is that the MAG canonical e.m. tensor is related  to the standard canonical e.m. tensor that one obtains from the Lagrangian ${\cal L}(\phi^A,\nabla_\m\phi^A;g)$ when it is coupled to Einstein gravity. This follows from
\beq
\label{tL}
\tau^{\mu}{}_{\nu}=\tau^{\mu}{}_{a}e_{\nu}{}^{a}
=-e^{\mu}{}_{c}\left\{\frac{1}{\sqrt{-g}}\frac{\partial \sqrt{-g}}{\partial e^{\nu}{}_{c}}\mathcal{L}_{e}+\frac{\partial \mathcal{L}_{e}}{\partial e^{\nu}{}_{c}  }  \right\}=\delta^{\mu}_{\nu}\mathcal{L}_{e}-e^{\mu}{}_{c}\frac{\partial \mathcal{L}_{e}}{\partial e^{\nu}{}_{c}  } \,,
\eeq
where we have used the identities
\beq
\frac{\partial e^{\lambda}{}_{c}}{\partial e_{\mu}{}^{a}}=-e^{\mu}{}_{c}e^{\lambda}{}_{a}\,,\quad
\frac{\partial \sqrt{-g}}{\partial e^{\nu}{}_{c}}=-\sqrt{-g}e_{\nu}{}^{c}\,.
\eeq
To take care of the last term on the right-hand side of (\ref{tL}) we must distinguish between the ordinary affine covariant derivatives and the MAG Weyl covariant derivatives. For ordinary affine covariant derivatives the vielbein appears only through the covariant derivative of the matter fields
\beq
\nabla_{a}\phi^A=e^{\rho}{}_{a}\nabla_{\rho}\phi^A\quad\Rightarrow\quad 
\frac{\partial (\nabla_{a}\phi^A)}{\partial e^{\lambda}{}_{c}}=\delta^{c}_{a}\nabla_{\lambda}\phi^A \label{derL}\,,
\eeq
since the spin connection is an independent variable. Then assuming that the matter action depends only algebraically on the vielbein we obtain
\beq
\frac{\partial \mathcal{L}_{e}}{\partial e^{\nu}{}_{c}  }=\frac{\partial\mathcal{L}_{e}}{\partial(\nabla_{a}\phi^A) }\frac{\partial(\nabla_{a}\phi^A)}{\partial e^{\nu}{}_{c}}=\frac{\partial\mathcal{L}_{e}}{\partial (\nabla_{c}\phi^A)}\nabla_{\nu}\phi^A\,.
\eeq
Inserting the above into \eqref{tL} we  obtain 
\beq
\label{tauordinary}
\tau^{\mu}{}_{\nu}
=-e^{\mu}{}_{c}\frac{\partial\mathcal{L}_{e}}{\partial \left(\nabla_{c}\phi^A\right) }\nabla_{\nu}\phi^A+\delta^{\mu}{}_{\nu}\mathcal{L}_{e}
=-\frac{\partial \mathcal{L}_{e}}{\partial (\nabla_{\mu}\phi^A)}\nabla_{\nu}\phi^A+\delta^{\mu}{}_{\nu}\mathcal{L}_{e}\equiv t^{\mu}{}_{\nu}\,.
\eeq
For Weyl covariant derivatives the vielbein appears also in the variation of the Weyl connection (i.e. torsion vector)~\cite{Iosifidis:2018zwo}, see also~\cite{Sauro:2022chz,Sauro:2022hoh}. For example consider the action of the Weyl covariant derivative on a tensor weight scalar
\beq
D_a\phi_A=e^\rho{}_a\,D_\rho\phi_A\,,\quad
D_\rho\phi_A=\nabla_\rho\phi_A-\alpha S_\rho\phi_A\,,\quad \alpha=\frac{2w_A}{d-1}\,,
\eeq
where $w_A$ is the weight of the field(s) $\phi_A$. The end result is
\begin{align}
\label{tauWeyl}
\tau_{(w)}^{\mu}{}_{\nu}&=-\frac{\partial \mathcal{L}_{e}}{\partial (D_{\mu}\phi^A)}D_{\nu}\phi^A+\delta^{\mu}{}_{\nu}\mathcal{L}_{e}+\alpha\hat\nabla_{\lambda}\left(\phi^{A}\frac{\partial \mathcal{L}_{e}}{\partial (D_{\rho}\phi^{A})}\delta^{[\mu}_{\rho}\delta^{\lambda]}_{\nu} \right)\nonumber\\
&= t^\m{}_{\n}+\alpha\hat\nabla_{\lambda}\left(\phi^{A}\frac{\partial \mathcal{L}_{e}}{\partial (D_{\rho}\phi^{A})}\delta^{[\mu}_{\rho}\delta^{\lambda]}_{\nu} \right)\,,
\end{align}
where we recall that $\hat\nabla_\mu=\nabla_\mu-2S_\mu$ and the subscript $w$ in $\tau_{(w)}^{\mu}{}_{\nu}$ denotes the inclusion of the Weyl covariant derivative. By construction, $\tau_{(w)}^{\mu}{}_{\nu}$ is traceless and the right-hand side of \eqref{tauWeyl} is the covariantized form of \eqref{dil2}.

Nevertheless, independently of the nature of the covariant derivatives, we can show that the tensors \eqref{cemt}, \eqref{grav.stress} and \eqref{hyper.def} are not independent. Using the chain rule and the identity \eqref{spin.constraint}, we find\footnote{The result (\ref{EMTs}) was originally found in Eq. (5.2.16) of \cite{Hehl:1994ue} expressed in exterior forms language and in a coordinate basis in Eq.(55) of \cite{Obukhov:2014nja} for $F=1$. See also \cite{Iosifidis:2020gth}. We review its derivation in Appendix \ref{Appendix.proof}. 
}
\beq
 \label{EMTs}
\tau^{\mu}{}_{\nu} = T^{\mu}{}_{\nu}+\frac{1}{2 \sqrt{-g}}\hat{\nabla}_\lambda(\sqrt{-g}\Delta_{\nu}{}^{\mu\lambda})\,.
\eeq
Eq. (\ref{EMTs}) tells us that the knowledge of $T^{\m\n}$ and $\Delta_\lambda{}^{\m\n}$ gives us access to the canonical e.m. tensor of the matter theory: in the case of ordinary affine covariant derivatives to the standard canonical e.m. tensor and in the case of a Weyl covariant derivatives to the {\it traceless part} of the standard canonical e.m. tensor (\ref{MAGtrace}).  This is a useful result in cases where the only information one has regarding the matter theory is the variation of its effective action wrt to the metric or other background fields, and hence there is no information regarding its canonical e.m. tensor. Such a case is, notably, AdS/CFT holography. Using  \eqref{EMTs} we further find  
\beq
\frac{1}{\sqrt{-g}}\hat{\nabla}_\mu(\sqrt{-g}\tau^{\mu}{}_{\alpha})=\frac{1}{2} \Delta^{\lambda\mu\nu}R_{\lambda\mu\nu\alpha}-\frac{1}{2}Q_{\alpha\mu\nu}T^{\mu\nu}-2 S_{\alpha\mu\nu}\tau^{\mu\nu}\,, \label{cc2}
\eeq
expressed in a holonomic frame  \cite{Hehl:1994ue,Iosifidis:2020gth}, see also~\cite{Lompay:2013mdd,Lompay:2013opa} for conservation laws in metric-torsion theories.

\subsection{The trivial flat limit and the improvement terms}

The simplest flat limit of the above results correspond to setting
the metric to be flat and the affine connection to be zero. We then obtain from \eqref{EMTs} and \eqref{cc2}  
\beq
\label{EMTs.flat}
\boxed{\tau^{\mu\nu}=T^{\mu\nu}+\frac{1}{2}\partial_{\lambda}\Delta^{\nu\mu\lambda}\,,\quad \partial_{\mu}\tau^{\mu\nu}=0\,.}
\eeq
Also, the flat limit of (\ref{MAGtrace}) reads 
\beq
\label{MAGtrace.flat}
\boxed{T^\mu{}_\mu+\frac12\partial_\lambda\Delta_\mu{}^{\mu\lambda}=0\,.}
\eeq
The results \eqref{EMTs.flat} and \eqref{MAGtrace.flat} correspond to the improvement process of the flat canonical e.m. tensor, and we see that the improvement terms are related to the flat limit of the hypermomentum. 

The flat limit of \eqref{conserved.j} yields the current
\beq
\label{jmu}
j^{\mu}
=-\xi_\lambda t^{\mu\lambda}+\frac{1}{2}\Delta^{\lambda\nu\mu}\partial_{\nu}\xi_{\lambda}\,,
\eeq
where $\xi^\m$ is an arbitrary vector field. This coincides with the corresponding result (\ref{cEq}) that was found from the Noether procedure in flat space, which clarifies further the role of the hypermomentum in the improvement process. 

Restricting then $\xi^\m$ to the Killing vectors of Minkowski, i.e. the translations and Lorentz generators, we will be able to show that the flat limit of the hypermomentum correspond to the BR improvent terms. From (\ref{jmu}) we obtain 
\beq
\label{sjdsksll}
\partial_\mu j^\mu=\frac12\left(-
2\partial_\mu\xi_\nu T^{\mu\nu}+\partial_\lambda\partial_\mu\xi_\nu\Delta^{\nu\lambda\mu}\right)\,,
\eeq
upon employing \eqref{EMTs.flat}. For translations $\xi^\mu=a^\mu$ the conservation of the current (\ref{jmu}) yields the on-shell conservation of the canonical energy--momentum tensor
\beq
\partial_\mu t^\mu{}_\nu=0\,.
\eeq
For Lorentz transformations $\xi^{\m}=\omega^{\m}{}_{\n}x^\n$ the current (\ref{jmu}) becomes 
\beq
\label{fksksaal}
(J^\mu)^{\nu\rho}=
t^{\mu\nu}x^\rho-t^{\mu\rho}x^\nu
-\Sigma^{\nu\rho\mu}\,,
\eeq
and its conservation gives 
\beq
\label{ckdfkjdk}
\partial_\mu(J^\mu)^{\nu\rho}=0\,\Rightarrow\,t^{\mu\nu}-t^{\nu\mu}+
\partial_\lambda\Sigma^{\mu\nu\lambda}=0\,,
\eeq
where  $\Sigma^{\mu\nu\lambda}$ is the spin part of the hypermomentum defined in \eqref{hyper.Spin}. 
We see that (\ref{ckdfkjdk}) is equivalent to the condition (\ref{Lorentz2}) if we identify
\beq
\label{PsiSigma}
\Sigma^{\m\n\rho}=-2\Psi^{\rho[\m\n]}
\eeq
and as promised the improvement terms used in (\ref{BR1}) correspond to the flat limit of the hypermomentum. This provides a geometric interpretation of the BR improvement procedure.

There is no improvement process for dilatations with  $\xi^\mu=\lambda\, x^\mu$. In flat space they just lead to the condition \eqref{dil2}  for the trace of the canonical e.m. tensor therefore we do not expect that the {\it virial current}  $N^\m$  in \eqref{dil2} is related to the hypermomentum. From \eqref{EMTs.flat} we obtain the corresponding trace condition for the symmetric e.m. tensor $T^{\m\n}$ as
\beq
\label{Ttrace}
T^{\m}{}_\m=\partial_\m\left(N^\m+\frac{1}{2}\Delta^\m\right)\,.
\eeq
The improvement procedure comes about when we consider special conformal transformations $\xi^\mu=b^\mu x^2-2x^\mu b_\nu x^\nu$ and 
the on-shell conserved tensor $\partial_\mu K^\mu{}_\nu=0$ reads
\beq
\begin{split}
\label{current.special.conformal}
K^\mu{}_\nu&=x^2t^\mu{}_\nu-2x_\nu x^\rho t^\mu{}_\rho
-2(x^\rho\Sigma_{\nu\rho}{}^\mu-x_\nu\Sigma^{\mu\rho}{}_\rho)
+2(x_\nu\partial_\rho Z^{\mu\rho}-Z^\mu{}_\nu)\\
&=x^2T_\text{BR}{}^\mu{}_\nu-2x_\nu x^\rho T_\text{BR}{}^\mu{}_\rho
+2(x_\nu\partial_\rho Z^{\mu\rho}-Z^\mu{}_\nu)
\,,
\end{split}
\eeq
provided that 
\beq
T_{\text{BR}\mu}{}^\mu=\partial_\mu\partial_\nu Z^{\mu\nu}\,,\quad Z^{\mu\nu}=Z^{\nu\mu}\,,
\eeq
which translates to the condition
\beq
\partial_\nu Z^{\mu\nu}=
\Lambda^\mu+\frac12\Delta^\mu-\Sigma^{\mu\nu}{}_\nu\,.
\eeq

\noindent
As we will see in detail in the examples that follow in Sections  \ref{rest.Lorentz}  and \ref{rest.conformal}, the r\^ole of the hypermomentum is the restoration of Lorentz  or scale invariance in the MAG set-up. The results will match those found using the BR or the virial current procedures.

\section{Lorentz invariance improvement terms}
We will present below explicit field theory examples where coupling to MAG correspond to the BR improvement process. 
\label{rest.Lorentz}

 \subsection{The electromagnetic field}
 \label{section.Maxwell}
The action for the free Maxwell field is
\beq
\label{Maxwell.action}
S_\text{EM}=\int 
d^dx\mathcal{L}_\text{EM}\,,\quad
\mathcal{L}_\text{EM}=-\frac{1}{4}F_{\mu\nu}F^{\mu\nu}\,,
\eeq
 where, as usual, $F_{\mu\nu}=\partial_{\mu}A_{\nu}-\partial_{\nu}A_{\mu}$  is the field strength of the electromagnetic field. The action is invariant under gauge transformations $A_\mu\mapsto A_\mu+\partial_\mu f$ and its equations of motion are given by
 \beq
 \label{EM.eom}
\partial_{\lambda}F^{\mu\lambda}=0\,.
 \eeq
 The corresponding canonical energy--momentum tensor is found to be
\beq
\label{canonicalEM}
t^{\mu\nu}=F^{\mu\lambda}\partial^{\nu}A_{\lambda}-\frac{1}{4}\eta^{\mu\nu}F_{\kappa\lambda}F^{\kappa\lambda}\,.
\eeq
This is neither symmetric nor gauge invariant. To remedy the situation and obtain an improved tensor that meets these requirements one uses the BR method, revisited in Appendix \ref{Maxwellapp}, which amounts to adding the term \cite{Landau:1975pou}
\beq
\label{Belin.extra}
-F^{\mu\lambda}\partial_{\lambda}A^{\nu}\,,
\eeq
giving the symmetric and also gauge invariant electromagnetic energy--momentum tensor
\beq
\label{stressMaxwell}
T^{\mu\nu}=F^{\mu\lambda}F^{\nu}{}_{\lambda}-\frac{1}{4}\eta^{\mu\nu}
F_{\kappa\lambda}F^{\kappa\lambda}\,.
\eeq
The e.m. tensor \eqref{stressMaxwell} can be alternatively obtained if we minimally couple the action \eqref{Maxwell.action} to Einstein gravity as
\beq
\label{MaxwellEinstein}
S_\text{EM}=-\frac{1}{4}\int  d^dx\sqrt{-g}\,g^{\m\rho}g^{\n\sigma}F_{\m\n}F_{\rho\sigma}
\eeq
and then use \eqref{grav.stress}. The resulting Hilbert metrical e.m. tensor coincides in the flat limit with \eqref{stressMaxwell}. However, as we have mentioned above the minimal coupling of the theory to Einstein gravity obscures the access to the canonical e.m. tensor. This is rectified by coupling the theory to MAG as follows. We consider a curved metric and replace
 \beq
 \label{hkddkjcd12}
\partial_{\mu} \mapsto \nabla_{\mu}
 \eeq
where $\nabla_{\mu}$ is now the covariant derivative with respect to a generic affine connection, then the field strength generalizes  to
\beq
\hat{F}_{\mu\nu}:=\nabla_\mu A_\nu-\nabla_\nu A_\mu=F_{\mu\nu}+2 S_{\mu\nu}{}{}^{\lambda}A_{\lambda}\,,
\eeq
which is not gauge invariant\footnote{The breaking of gauge invariance 
\begin{equation*}
\delta_\Lambda\hat{\cal L}_\text{EM}=-S_{\mu\nu}{}^\rho\hat F^{\mu\nu}\partial_\rho \Lambda\,,\quad \delta_\Lambda A_\mu=\partial_\mu\Lambda\,,    
\end{equation*}
is not fatal at this point as this is only an intermediate step as at the end we will go back to the flat background (i.e. we set $g=\eta$ and $\Gamma^{\lambda}{}_{\mu\nu}=0$) and the gauge invariance will then be restored.} and 
the corresponding Lagrangian becomes
\beq
\label{fheksdls}
\begin{split}
\hat{\mathcal{L}}_\text{EM}&=-\frac{1}{4}\hat{F}_{\mu\nu}\hat{F}^{\mu\nu}\\
&=-\frac{1}{4}\left(F_{\mu\nu}F^{\mu\nu}+4S_{\mu\nu}{}^\rho F^{\mu\nu}A_\rho+4S_{\mu\nu}{}^\rho S^{\mu\nu\lambda}A_\rho A_\lambda\right)\,.
\end{split}
\eeq
The associated hypermomentum to this matter sector is easily computed using \eqref{hyper.def}, yielding in the flat spacetime limit
\beq
\label{nxhkqo340}
\Delta_{\lambda}{}^{\mu\nu}=
2 A_{\lambda} F^{\mu\nu}\,.
\eeq
Then, using the equations of motion \eqref{EM.eom} we obtain
\beq
\frac{1}{2}\partial_{\lambda}\Delta^{\nu\mu\lambda}= F^{\mu\lambda}\partial_{\lambda}A^{\nu}\,,
\eeq
which demonstrates that the hypermomentum provides a geometric origin for  the BR improvement term \eqref{Belin.extra}. The above discussion can be trivially extended for the case of a Proca field.

A comment is in order concerning the scale invariance of electromagnetism \eqref{Maxwell.action} but non conformal invariance in $d\neq4$.
Indeed by taking the trace of \eqref{stressMaxwell} we find that it takes on-shell the form of a total derivative
\beq
\label{bjkssl}
T^\mu{}_\mu=-\frac{d-4}{4}F_{\mu\nu}F^{\mu\nu}=\partial_\mu V^\mu\,,\quad 
V^\mu=-\frac{d-4}{2}A_\nu F^{\mu\nu}\,.
\eeq
From the above and equation \eqref{skkn1} we can construct an on-shell conserved dilation current. Despite been non-gauge invariant, the corresponding dilation charge $Q=\int d^{d-1}x\, j^0_D\,,$ is on-shell gauge invariant up to boundary terms.  
However, in this case the scale does not implement conformal invariance. More concretely, $V^\mu$ in \eqref{bjkssl} cannot be expressed as  $\partial_\nu Z^{\mu\nu}$, where $Z^{\mu\nu}$ is a symmetric rank-2 tensor, while keeping gauge invariance.\footnote{In detail, there exists such a rank-2 tensor, namely
\begin{equation*}
Z^{\mu\nu}=-\frac{d-4}{2}\left(A^\mu A^\nu-\frac12\,\eta^{\mu\nu} A_\kappa A^\kappa\right)\,,  
\end{equation*}
upon enforcing the Lorentz gauge $\partial_\mu A^\mu=0$. However, the corresponding symmetric, conserved and traceless tensor $\Theta^{\mu\nu}$ read through \eqref{sconf5} and \eqref{stressMaxwell} as well the corresponding conserved charges defined through \eqref{currents.scale.special}, e.g.
$
Q_\nu=\int d^{d-1}x\, K^0{}_\nu\,,
$
explicitly depend on the gauge field. Hence, scale does not extend to conformal invariance without breaking the gauge symmetry of the theory.}

\subsubsection{Abelian $p$-forms}

The above easily extends to the action of Abelian $p$-forms
\beq
\label{Lp}
{\cal L}_p=-\frac{1}{2p!} F_{\mu_1\mu_2\dots\mu_p}F^{\mu_1\mu_2\dots\mu_p}\,,\quad 
F_{\mu_1\mu_2\dots\mu_p}=p\,\partial_{[\mu_1} A_{\mu_2\dots\mu_p]}\,,
\eeq
whose equations of motion read
\beq
\partial_{\mu_1}F^{\mu_1\mu_2\dots\mu_p}=0\,,
\eeq 
where $A_{\mu_2\dots\mu_{p}}$ is a massless $p-1$ form.
The corresponding canonical e.m. tensor is then given by
\beq
\label{emDp}
t^{\mu\nu}=\frac{1}{(p-1)!}F^{\mu\mu_2\dots\mu_p}\partial^\nu A_{\mu_2\dots\mu_p}-\frac{1}{2p!}\eta^{\mu\nu}F_{\mu_1\mu_2\dots\mu_p}F^{\mu_1\mu_2\dots\mu_p}\,,
\eeq
which is neither symmetric nor gauge invariant under the following transformation
\beq
A_{\mu_1\dots\mu_{p-1} }\mapsto A_{\mu_1\dots\mu_{p-1}}+(p-1)\partial_{[\mu_1}\Lambda_{\mu_2\dots\mu_{p-1}]}\,,
\eeq
with $\Lambda_{\mu_2\dots\mu_{p-1}}$ a $p-2$ form. Similarly to electromagnetism the BR improvement of \eqref{emDp} comes from the hypermomentum. For that we just need the minimal coupling of \eqref{Lp} to MAG according \eqref{hkddkjcd12}, and we find
\beq
\hat F_{\mu_1\mu_2\dots\mu_p}=p\,\nabla_{[\mu_1} A_{\mu_2\dots\mu_p]}=F_{\mu_1\mu_2\dots\mu_p}+pS_{[\mu_1\mu_2}{}^\rho
A_{|\rho|\mu_3\dots\mu_p]}\,,
\eeq
where $\rho$ is excluded from the antisymmetry. The corresponding action is given by
\beq
\hat {\cal L}_p=-\frac{1}{2p!}\hat F_{\mu_1\mu_2\dots\mu_p}\hat F^{\mu_1\mu_2\dots\mu_p}\,,
\eeq
whereas the associated hypermomentum can be computed through \eqref{hyper.def}
\beq
\Delta_\lambda{}^{\mu\nu}=\frac{2}{(p-2)!}A_{\lambda\mu_3\dots\mu_p}F^{\mu\nu\mu_3\dots\mu_p}\,.
\eeq
Inserting the latter and \eqref{emDp} into \eqref{EMTs.flat} we find on-shell
\beq
T^{\mu\nu}=\frac{1}{(p-1)!}\left(F^{\mu\mu_2\dots\mu_p}F^\nu{}_{\mu_2\dots\mu_p}-\frac{1}{2p}\eta^{\mu\nu}F_{\mu_1\mu_2\dots\mu_p}F^{\mu_1\mu_2\dots\mu_p}\right)\,,
\eeq
which is symmetric and gauge invariant. Its trace takes on-shell  the form of a total derivative
\beq
T^\mu{}_\mu=-\frac{d-2p}{2p!}F_{\mu_1\mu_2\dots\mu_p}F^{\mu_1\mu_2\dots\mu_p}=\partial_\mu\Lambda^\mu\,,\quad 
\Lambda^\mu=-\frac{d-2p}{2(p-1)!} A_{\mu_2\dots\mu_p} F^{\mu\mu_2\dots\mu_p}\,.
\eeq
Extending the discussion after \eqref{bjkssl}, we find that the theory is scale but non-conformal invariant for $d\neq 2p$.

\subsection{The Dirac field }

Let us consider the massive Dirac field in a curved $d$ dimensional spacetime with a mostly plus metric described in terms of the Hermitian action\footnote{The gamma matrices in the mostly minus signature metrics are related to those of the mostly plus (we adopt here) as follows, $\gamma^\mu\big{|}_\text{mostly minus}=i\,\gamma^\mu\big{|}_\text{mostly plus}$.}
\beq
\label{Dirac.action.curv}
S_\text{D}=\int d^dx\sqrt{-g} {\cal L}_\text{D}\,,\quad
{\cal L}_\text{D}=-\frac{i}{2}\left(\bar\psi\gamma^c\nabla_c\psi-\nabla_c\bar\psi\gamma^c\psi\right)-im\bar\psi\psi\,,
\eeq
where $\bar\psi=\psi^\dagger\gamma^0$.
The $\gamma$-matrices obey the Clifford algebra
\beq
\label{Clifford.alg}
\gamma^{a}\gamma^{b}+\gamma^{b}\gamma^{a}=2 \eta^{ab}\,,
\eeq
where $(\gamma^0)^\dagger=-\gamma^0$ and $(\gamma^i)^\dagger=\gamma^i$ or simply $(\gamma^a)^\dagger=\gamma^0\gamma^a\gamma^0$.
In addition, the $\gamma$'s obey the useful identity
\beq
\label{jdjsklds}
\begin{split}
&[\gamma^a,\Sigma^{bc}]
=\eta^{ab}\gamma^c-
\eta^{ac}\gamma^b\,,
\end{split}
\eeq
where $\Sigma^{ab}=\frac14[\gamma^a,\gamma^b]$.
In \eqref{Dirac.action.curv} the torsionfull covariant derivatives act on the spinor fields $\psi$ and $\bar\psi$ on the tangent frame metric as (see 8.35 of \cite{SUGRA}).
\beq
\nabla_c\psi=\partial_c\psi+\frac12\omega_{abc}\Sigma^{ab}\psi \,,\quad
\nabla_c\bar\psi=\partial_c\bar\psi-\frac12\omega_{abc}\bar\psi\Sigma^{ab}\,,
\label{DiracTor}
\eeq
where $\partial_c=e^\mu{}_c\partial_\mu$ and 
$\omega_{abc}$ are the spin connection coefficients, being antisymmetric in $a,b$ indices.\footnote{This is so by demanding metricity 
$Q_{\rho\mu\nu}=-\nabla_{\rho}g_{\mu\nu}=0$ in the case at hand. For spaces with non-metricity, the generalization to the Dirac equation is not known. We thank Friedrich Hehl for related discussions on this point.} The spinor field transforms under infinitesimal Lorentz transformations \eqref{Lorentz.transf}
where
\beq
\label{Lorentz.transf.spinor}
\delta_\omega\psi=-\frac12\omega_{ab}\Sigma^{ab}\psi\,,\quad
\delta_\omega\bar\psi=\frac12\omega_{ab}\bar\psi
\Sigma^{ab}\,,
\eeq
where $\omega_{ab}$ are the infinitesimal parameters not to be confused with the spin-connection coefficients $\omega_{abc}$.
We can also derive the equations of motion of \eqref{Dirac.action.curv}. It is a rather long but straightforward calculation to show that
\beq
\label{eomDirac}
\gamma^a\left(\nabla_a-S_a\right)\psi+m\psi=0\,,\quad
(\nabla_a-S_a)\bar\psi \gamma^a-m\bar\psi=0\,,
\eeq
that is in agreement with~\cite{Hehl.Datta}. 

\noindent
Starting from the action \eqref{Dirac.action.curv} we can compute the hypermomentum  \eqref{hyper.def} which in the flat spacetime takes the form
\beq
\label{Hyper.Dirac}
\Delta^{\nu\mu\lambda}=-\frac{i}{2}\bar\psi\left(\gamma^\lambda\Sigma^{\mu\nu}+\Sigma^{\mu\nu}\gamma^\lambda\right)\psi\,,
\eeq
where we have used \eqref{spin.constraint}. In addition, using the Clifford algebra \eqref{Clifford.alg} we find that the hypermomentum tensor turns out to be totally antisymmetric (in agreement with~\cite{Hehl.Datta}) since
\beq
\label{Hyper.Dirac1}
\Delta^{\nu\mu\lambda}=-i\bar\psi\gamma^{[\lambda}\Sigma^{\mu\nu]}\psi=\frac{i}{2}\bar\psi \gamma^{[\nu}\gamma^\mu\gamma^{\lambda]}\psi\,.
\eeq
We can easily find the divergence of \eqref{hyper.def}
\beq
\label{dfjskdjsk}
\partial_\lambda \Delta^{\nu\mu\lambda}=i\bar\psi\left(\gamma^\mu\partial^\nu-\gamma^\nu\partial^\mu\right)\psi+\frac{i}{2}\partial^\mu\left(\bar\psi\gamma^\nu\psi\right)-
\frac{i}{2}\partial^\nu\left(\bar\psi\gamma^\mu\psi\right)\,,
\eeq
where we used the equations of motion \eqref{eomDirac} in flat spacetime 
and the third of \eqref{jdjsklds}. Now, the corresponding canonical on-shell  energy--momentum tensor of the Dirac field is derived in Appendix~\ref{Diracapp} in equation \eqref{vdkdsksk} and it is restated here for reader's convenience
\beq
\label{vdkdsksk0}
t^{\mu\nu}=
\frac{i}{2}\left(\bar\psi\gamma^\mu\partial^\nu\psi-\partial^\nu\bar\psi\gamma^\mu\psi\right)\,.
\eeq
Substituting \eqref{vdkdsksk0} and \eqref{dfjskdjsk} into \eqref{EMTs.flat} we find the conserved improved tensor (also derived in~\cite{Hehl:1976vr})
\beq
\label{improved.mostly.plus}
T^{\mu\nu}=\frac{i}{4}
\left(\bar\psi\gamma^\mu\partial^\nu\psi-
\partial^\nu\bar\psi\gamma^\mu\psi\right)
+
\frac{i}{4}
\left(\bar\psi\gamma^\nu\partial^\mu\psi-
\partial^\mu\bar\psi\gamma^\nu\psi\right)\,,
\eeq
where we note $T^\mu{}_\mu=-im\bar\psi\psi$. This coincides with the stress energy tensor obtained using the BR method in Appendix~\ref{Diracapp} and equation \eqref{BR.Dirac}. From \eqref{Hyper.Dirac1} and \eqref{EMTs.flat} we also see explicitly why the usual treatment of taking the symmetric part of $t_{\mu\nu}$ in \eqref{vdkdsksk0} for the Dirac field, in order to derive the symmetric energy--momentum tensor, works in this case. The reason is that the hypermomentum is totally antisymmetric \eqref{Hyper.Dirac1} and thus the symmetric part of $t_{\mu\nu}$ in \eqref{EMTs.flat} equals to
\beq
t_{(\mu\nu)}=T_{\mu\nu}
\eeq
and also $T_{\mu\nu}$ is conserved. In our formulation, therefore, it becomes clear why this method of simply symmetrizing the canonical stress energy tensor \eqref{vdkdsksk0} works out in the case of the Dirac field. Lastly the gravitational definition of the energy--momentum tensor \eqref{grav.stress} fails here since spinors do not couple to the metric but rather to the vielbeins. This also proves the superiority of our method over the standard procedure, since in our approach no metric variations are performed.

\subsection{Higher derivative scalar theories}

We now move to a nontrivial example of the Belinfante--Rosenfeld improvement process which demonstrates the power of our approach. 
Consider higher-derivative scalar theories as those considered e.g. in~\cite{Osborn:2016bev} with action
\beq
\label{dsnksdw}
S_4=-\frac12\int d^d x\, (\partial^2\phi)^2\,,\quad \partial^2\phi=\eta^{\mu\nu}\partial_\mu\partial_\nu\phi\,, 
\eeq
whose equations of motion are given by
\beq
\label{eom.4phi}
\partial^2\partial^2\phi=0\,.
\eeq
Since the Lagrangian in \eqref{dsnksdw} is higher derivative the the canonical e.m. tensor is given by 
\begin{align}
\label{Noether.general}
t^{\mu\nu}&=-\frac{\partial \mathcal{L}}{\partial (\partial_\mu\phi)}\partial^\nu\phi+\eta^{\mu\nu}\mathcal{L}-2\frac{\partial{\cal L}}{\partial(\partial_\mu\partial_\lambda\phi)}\partial^\nu\partial_\lambda\phi+
\partial_\lambda\left(\frac{\partial{\cal L}}{\partial(\partial_\mu\partial_\lambda\phi)}\partial^\nu\phi\right)\,,\nonumber \\
&=2\partial^\mu\partial^\nu\phi\partial^2\phi-\frac12\eta^{\mu\nu}\,(\partial^2\phi)^2-\partial^\mu\left(\partial^2\phi\partial^\nu\phi\right)\,.
\end{align}
This can be improved (see Appendix~\ref{nonunitaryapp} for details) to the following symmetric BR e.m. tensor 
\beq
\label{qpds12tu}
\begin{split}
T_\text{BR}^{\mu\nu}&=2\partial^\mu\partial^\nu\phi\partial^2\phi-\frac12\eta^{\mu\nu}\,(\partial^2\phi)^2-\partial^\mu\left(\partial^2\phi\partial^\nu\phi\right)-\partial^\nu(\partial^2\phi\partial^\mu\phi)+\eta^{\mu\nu}\partial_\lambda(\partial^2\phi\partial^\lambda\phi)\,,
\end{split}
\eeq
where last two terms correspond to the BR improvement terms. 

We can then show that the above BR e.m. tensor coincides with the flat limit of the following Hilbert metrical e.m. tensor 
\beq
\label{nbdk1s2kw}
\begin{split}
T_\text{BR}^{\mu\nu}&=2\tilde\nabla^\mu\partial^\nu\phi\tilde\nabla^2\phi-\frac12\eta^{\mu\nu}\,(\tilde\nabla^2\phi)^2-\tilde\nabla^\mu\left(\tilde\nabla^2\phi\tilde\partial^\nu\phi\right)-\tilde\nabla^\nu(\tilde\nabla^2\phi\partial^\mu\phi)+\eta^{\mu\nu}\tilde\nabla_\lambda(\partial^2\phi\partial^\lambda\phi)\,,
\end{split}
\eeq
which is in turn obtained using \eqref{grav.stress} when we couple minimally couple the action \eqref{dsnksdw} to Einstein gravity.  As before, the latter procedure does not give us a geometric understanding of the BR improvement terms. For that we would need access to the canonical e.m. tensor and this can be achieved if we couple the theory to MAG using the following
\beq
\label{dsnksdw1}
S_4=-\frac12\int d^d x\sqrt{-g}\, (\nabla^2\phi)^2\,,\quad \nabla^2\phi=g^{\mu\nu}\nabla_\mu\partial_\nu\phi\,,
\eeq
where now $\nabla_\m$ is the affine connection. The crucial point is that when we use 
\eqref{grav.stress} to compute the Hilbert metrical e.m. tensor from \eqref{dsnksdw1} we need to take into account the nontrivial connection and the result is not the same as (\ref{nbdk1s2kw}). In fact the corresponding e.m. tensor is not even covariantly conserved! Restricting to flat space we obtain 
\beq
\label{sksllskaq}
T^{\mu\nu}=2\partial^\mu\partial^\nu\phi\partial^2\phi-\frac12\eta^{\mu\nu}\,\partial^2\phi \partial^2\phi\,.
\eeq
The reason that \eqref{sksllskaq} is not conserved is the presence of hypermomentum which we can calculate using \eqref{hyper.def} and the definition of the covariant Laplacian
\beq
\label{sksns}
\nabla^2\phi=
g^{\mu\nu}\left(\partial_\mu\partial_\nu\phi-\Gamma_{\nu\mu}{}^\rho\partial_\rho \phi\right),
\eeq
as
\beq
\label{sksns1}
\Delta^{\nu\mu\lambda}=-2\eta^{\mu\lambda}\partial^2\phi\partial^\nu\phi\,.
\eeq
We can verify that using \eqref{sksllskaq} and \eqref{sksns1} into \eqref{EMTs.flat} we find the canonical e.m. tensor \eqref{Noether.general}. 

Although the MAG metrical e.m. tensor \eqref{sksllskaq} does not coincide with the BR one, we can still show that the improvement terms in \eqref{qpds12tu} come from the hypermomentum. For that we can use \eqref{Lorentz2} and \eqref{EMTs.flat} to establish that 
\beq
\label{PsiDelta}
\Psi^{\rho[\m\n]}=\frac{1}{2}\Delta^{[\n\m]\rho}\,.
\eeq
Then, from \eqref{BR1} we find that the BR improvement terms are generally given by
\beq
\label{improveDelta}
\frac{1}{2}\partial_\lambda\left(\Delta^{[\lambda\n]\m}+\Delta^{[\lambda\m]\n}+\Delta^{[\n\m]\lambda}\right)=\partial^\n(\partial^2\phi\partial^\m\phi)-\eta^{\m\n}\partial_\lambda(\partial^2\phi\partial^\lambda\phi)\,,
\eeq
which are exactly minus the improvement terms of \eqref{qpds12tu}, verifying our claim.

\section{Special conformal invariance improvement terms}
\label{rest.conformal}

We present below examples where coupling to MAG reveals the relevance of hypermomentum to the improvement process of the e.m. tensor in a conformal field theory.

\subsection{The massless scalar}

The simplest example is that of of massless free scalar field in $d$-dimensions with action
\beq
\label{action.free}
S_\text{scalar}=-\frac12\int d^dx\,(\partial \phi)^{2}\,,\quad (\partial \phi)^{2}=\eta^{\kappa\lambda}\partial_\kappa\phi\partial_\lambda\phi\,.
\eeq
The canonical e.m. tensor of (\ref{action.free}) is symmetric and coincides with the improved BR one as
\beq
\label{emconfscalar}
T^{\mu\nu}=\partial^\mu\phi \partial^\nu\phi -\frac12\eta^{\mu\nu}(\partial \phi)^{2}\,.
\eeq
This is the e.m. tensor that we would have obtained as the flat limit of the metrical e.m. tensor when we minimally couple the action \eqref{action.free} to Einstein gravity. However, the action \eqref{action.free} is scale invariant and this gives rise to an on-shell conserved dilatation current as 
\beq
  \label{skkn1}
 j^\mu_D=
 x_\nu T^{\mu\nu}-V^\mu\,.
 \eeq
The e.m. tensor \eqref{emconfscalar} has non-vanishing trace which takes the form of a total derivative
\beq
T^\mu{}_\mu=-\frac{d-2}{2}(\partial\phi)^2
 =\partial_\mu V^\mu\,,\quad 
 V^\mu=-\frac{d-2}{4}\partial^\mu\phi^2\,.
 \label{skkn0}
 \eeq
Moreover, since 
\beq
\label{skkn20}
V^\mu=\partial_\nu Z^{\mu\nu}\,,\quad
Z^{\mu\nu}=-\frac{d-2}{4}\eta^{\mu\nu}\phi^2\,,
\eeq
condition \eqref{sconf4} holds and using \eqref{sconf5} we can build an improved traceless e.m. tensor~\cite{Gross:1970tb}
\beq
\label{skkn}
\Theta_{\mu\nu}= \partial_{\mu}\phi \partial_{\nu}\phi -\frac{1}{2}\eta_{\mu\nu}(\partial \phi)^{2} 
+\frac{1}{4}\frac{d-2}{d-1}\left(\eta_{\mu\nu} \partial^2\phi^2-\partial_\mu\partial_\nu\phi^2\right)\,,
\eeq
which is conserved, symmetric and also traceless. The corresponding scale and special conformal currents can be read through \eqref{currents.scale.special}.
Equivalently we find the same expression using the improved energy--momentum tensor
\beq
\label{skknwww}
T_\text{imp}^{\mu\nu}= t^{\mu\nu}-\frac{1}{d-1}\left(\eta^{\mu\nu}\partial_\kappa V^\kappa -\partial^\nu V^\mu\right)\,.
\eeq
For $d=4$ it coincides with the improved stress energy tensor in~\cite{Callar.Jackiw} which  
was also derived in Appendix A of~\cite{Hill:2014mqa} by adding the total divergence term $\partial^2\phi$ with an appropriate coefficient in the action~\eqref{action.free}.

The usual way to find the improved conformal e.m. tensor $\Theta^{\m\n}$ is to couple the matter action \eqref{action.free} to Einstein gravity in a Weyl invariant way and this corresponds to finding the Weyl invariant extension of the flat Laplacian~\cite{Callar.Jackiw}. That process is well-known but it does not provide a geometric origin for the improvement terms in \eqref{skkn}. Here we will show that coupling the action \eqref{action.free} to MAG and requiring an extended Weyl invariance, namely local frame invariance, is also straightforward and does give a geometric meaning to the improvement terms through hypermomentum. The frame invariant extension of \eqref{action.free} can be achieved if we used the extended covariant derivative that  takes the form of a minimal coupling with a Weyl connection~\cite{Folland,Hall,Ciambelli:2019bzz}, see also~\cite{Iosifidis:2018zwo} and~\cite{Sauro:2022chz,Sauro:2022hoh} in the context of MAG. Following the latter reference we define 
\beq
\label{cov.Weyl}
\partial_\m\rightarrow \nabla_\m\rightarrow D_{\mu}=\nabla_\m+\alpha S_\m\,.
\eeq
The coefficient $\alpha$ will be found requiring that the resulting action 
\beq
\label{djckdks}
S=-\frac12\int d^dx\sqrt{-g}\,(D \phi)^{2}\,,
\eeq
is invariant under the local rescaling 
\beq
\label{skqseqs}
 \phi\mapsto e^{w\Omega}\phi\,,\quad
 w=\frac{d-2}{2}\,,
\eeq
accompanied by the frame rescaling \eqref{connection.Weyl}. From the latter we find that the torsion tensor transforms as
\beq
\label{frame.torsion}
S_{\mu\nu}{}^\rho\mapsto
S_{\mu\nu}{}^\rho
+\frac{1}{2}\left(\partial_\mu\Omega\delta_\nu^\rho-\partial_\nu\Omega\delta_\mu^\rho\right)
\eeq
and the torsion vector  \eqref{torsion2} as 
\beq
\label{frame.torsion.vector}
S_\mu\mapsto S_\mu+\frac{d-1}{2}\partial_{\mu}\Omega\,.
\eeq
We then find that for 
\beq
\alpha=\frac{d-2}{d-1}\,,
\eeq
the Weyl covariant derivative \eqref{cov.Weyl} transforms simply as
\beq
D_\mu\phi\mapsto e^{\frac{d-2}{2}\Omega} D_\mu\phi\,,
\eeq
yielding an invariant action \eqref{djckdks}.  Note also that  the Ricci scalar $\tilde R$ transforms as
\beq
\label{Weyl.Ricci}
\tilde R
\mapsto
e^{2\Omega}\left(\tilde R+2(d-1)\tilde\nabla^2\Omega-(d-1)(d-2)(\partial\Omega)^2\right)\,.
\eeq

We can now compute the hypermomentum \eqref{hyper.def} which in the flat limit becomes
 \beq
\Delta_{\nu}^{\;\;\mu\lambda}=\frac{d-2}{d-1} \delta^{[\mu}_{\nu}\partial^{\lambda]}\phi^2\,.
\label{hcskdshsksk}
\eeq
This can be decomposed as in \eqref{hypsplit}
\beq
\begin{split}
&\Delta^{\nu\mu\lambda}=
\Sigma^{\nu\mu\lambda}+\frac{\eta^{\mu\nu}}{d}\Delta^\lambda+\hat\Delta^{\nu\mu\lambda}\,,\quad\Sigma^{\nu\mu\lambda}=
\frac12\frac{d-2}{d-1}\eta^{\lambda[\mu}\partial^{\nu]}\phi^2\,,\\
&
\Delta^\lambda=\frac{d-2}{2}\partial^\lambda\phi^2\,,\quad
\hat\Delta^{\nu\mu\lambda}=\frac12\frac{d-2}{d-1}\left(\frac{\eta^{\mu\nu}}{d}\partial^\lambda-\eta^{\lambda(\mu}\partial^{\nu)}\right)\phi^2\,.
\end{split}
\eeq
Then, starting from \eqref{hcskdshsksk} we find
\beq
\partial_\lambda\Delta^{\nu\mu\lambda}=
\frac{1}{d}\eta^{\m\n}\partial_\lambda\Delta^\lambda+\partial_\lambda\hat\Delta^{\nu\mu\lambda}=
\frac{d-2}{2(d-1)}\left(\eta^{\mu\nu} \partial^2-\partial^\mu\partial^\nu\right)\phi^2\,,
\label{D}
\eeq
where $\partial_\lambda\Sigma^{\nu\mu\lambda}=0$.
Substituting  \eqref{D} into \eqref{EMTs.flat} we finally find (also derived in~\cite{Hehl:1976vr})
\beq
t^{\mu\nu}=\partial^{\mu}\phi \partial^{\nu}\phi -\frac{1}{2}\eta^{\mu\nu}(\partial \phi)^{2} 
+\frac{1}{4}\frac{d-2}{d-1}\left(\eta^{\mu\nu} \partial^2\phi^2-\partial^\mu\partial^\nu\phi^2\right)\,,
\eeq
which coincides with \eqref{skkn}.

 In fact we will show that in teleparallel gravity (see~\cite{Aldrovandi:2013wha} for a review), where we consider connections with vanishing curvature for the connection $\Gamma^{\lambda}{}_{\mu\nu}$, the action \eqref{djckdks} is equivalent with that of the conformally coupled scalar given in Eq. \eqref{action.conf} of Appendix \ref{Conf.Appendix}, up to a boundary term. To prove this we consider the Ricci scalar $R$ of the MAG theory, e.g. Eq.(13) in~\cite{Iosifidis:2021tvx}, specialized to vanishing non-metricity\footnote{We note that under a frame rescaling the Ricci scalar simply transforms as $R\mapsto e^{2\Omega} R$, where the additional pieces  in the transformations of the torsion, the torsion vector and the $\tilde R$, that is Eqs.\eqref{frame.torsion}, \eqref{frame.torsion.vector} and \eqref{Weyl.Ricci}  drop out.} 
\begin{equation}
\label{jfhdjslsq}
\begin{split}
R&=\tilde R+S_{\mu\nu\rho}S^{\mu\nu\rho}-2S_{\mu\nu\rho}S^{\rho\mu\nu}-4S_\mu S^\mu-4\tilde\nabla_\mu S^\mu\\
&=\tilde R-4\frac{d-2}{d-1}S_\mu S^\mu-4\tilde\nabla_\mu S^\mu\,,
\end{split}
\end{equation}
where in the second equality we have assumed  a vectorial type torsion of the form
\beq
S_{\mu\nu}{}^\rho=
\frac{1}{d-1}(S_\mu\delta_\nu^\rho-S_\nu\delta_\mu^\rho)\,.\label{vector}
\eeq
Then, we consider 
 \eqref{djckdks} which can be rewritten as
\begin{equation}
\label{nksdsksks}
S=\int d^dx\sqrt{-g}\left\{-\frac{1}{2}(\partial\phi)^2-\frac12\frac{d-2}{d-1}\left(\frac{d-2}{d-1}S_\mu S^\mu \phi^2-S^\mu\partial_\mu\phi^2\right)\right\}\,,
\end{equation}
or upon using equation \eqref{jfhdjslsq} we find
\begin{equation}
\label{nksdsksks1}
S=\int d^dx\sqrt{-g}\left\{-\frac{1}{2}(\partial\phi)^2-\frac12\frac{d-2}{d-1}\left(\frac14(\tilde R-R)\phi^2-\tilde\nabla_\mu(S^\mu\phi^2)\right)\right\}\,.
\end{equation}
Ignoring the last term (as it is a boundary term) and imposing flatness of the connection i.e. $R^{\lambda}{}_{\alpha\beta\gamma}=0$ we have  $R=0$ and we find exact agreement with the action of the well known action of the conformally coupled scalar \cite{Callar.Jackiw}
\beq
\label{action.conf}
S_\text{cs}=\int d^dx\sqrt{-g}\left(-\frac12(\partial\phi)^2-\frac18\frac{d-2}{d-1} \tilde R\phi^2\right)\,,
\eeq
which is revisited in Appendix~\ref{Conf.Appendix}.

\subsection{Higher derivative scalars}
\label{Sec.nonunitary}

The action \eqref{dsnksdw} is scale invariant. The trace of the canonical and the BR e.m. tensors \eqref{Noether.general} and \eqref{qpds12tu}  respectively, take on-shell respectively the forms
\beq
\label{sjcslal}
t^\mu{}_\mu=
\partial_\mu\tilde V^\mu\,,\quad
T_\text{BR}^\mu{}_\mu=
\partial_\mu V^\mu\,,
\eeq
where the virial currents are given by
\begin{align}
\tilde V^\mu&=
\frac12\left((2-d)\partial^\mu\phi\partial^2\phi+(d-4)\phi\partial^\mu\partial^2\phi\right)
\,,\\ 
V^\mu&=
\frac12\left(d\partial^\mu\phi\partial^2\phi+(d-4)\phi\partial^\mu\partial^2\phi\right)\,.
\end{align}
The corresponding conserved dilatation current takes the form of Eq. \eqref{skkn1}. Moreover, the theory is also conformal invariant \cite{Polchinski:1987dy,Conformal1} since the virial currents take the form since 
\beq
\label{sjcslal1}
\tilde V^\mu=\partial_\nu \tilde Z^{\mu\nu}\,,\quad
 V^\mu=\partial_\nu  Z^{\mu\nu}\,,
\eeq
with
\beq
\label{Zs.non.unitary}
\begin{split}
&\tilde Z^{\mu\nu}=-\frac{d-3}{2}\left(2\partial^\mu\phi\partial^\nu\phi-\eta^{\mu\nu}(\partial\phi)^2\right)+\frac12(d-4)\eta^{\mu\nu}\phi\partial^2\phi\,,
\\
& Z^{\mu\nu}=2\partial^\mu\phi\partial^\nu\phi-\eta^{\mu\nu}(\partial\phi)^2+\frac12(d-4)\eta^{\mu\nu}\phi\partial^2\phi\,.
\end{split}
\eeq
From the latter we can construct the conserved Noether tensor \eqref{current.special.conformal}
\beq
\partial_\mu K^{\mu\nu}=0\,,\quad K^{\mu\nu}=x^2T_\text{BR}^{\mu\nu}-2x^\nu x_\kappa T_\text{BR}^{\mu\kappa}+2(\partial_\kappa Z^{\mu\kappa}x^\nu-Z^{\mu\nu})\,.
\eeq
Having at hand the BR improved e.m. tensor, as well as $Z^{\m\n}$ we can use formula \eqref{sconf5} to construct the conformally improved, conserved symmetric and traceless e.m. tensor $\Theta^{\m\n}$ as (see also Eqs.(4.4) and (4.5) in \cite{Osborn.CFTs})
\beq
\begin{split}
\label{Thetancs}
\Theta_{\m\n}&=2\partial_\mu\partial_\nu\phi\partial^2\phi-\frac12\eta_{\mu\nu}\,(\partial^2\phi)^2 -\partial_\mu\left(\partial^2\phi\partial_\nu\phi\right)\\
&-\partial_\nu\left(\partial^2\phi\partial_\mu\phi\right)+\eta_{\mu\nu}\partial_\lambda(\partial^2\phi\partial^\lambda\phi)\\
&+\frac12\frac{d-4}{d-1}\left(\partial_\mu\partial_\nu-\eta_{\mu\nu}\partial^2\right)(\phi\partial^2\phi)\\
&+\frac{2}{d-2}\left(2\partial_\rho(\partial_\mu\partial_\nu\phi\partial^\rho\phi)-\eta_{\mu\nu}\partial_\rho\partial_\sigma(\partial^\rho\phi\partial^\sigma\phi)\right)\\
&-\frac{d}{(d-1)(d-2)}\left(\partial_\mu\partial_\nu-\eta_{\mu\nu}\partial^2\right)(\partial\phi)^2\,,
\end{split}
\eeq
obeying
\beq
\partial^\mu \Theta_{\m\n}=-\partial_\nu\phi\,\partial^2\partial^2\phi\,,\quad \eta^{\mu\nu}\Theta_{\m\n}=-\frac{1}{2}(d-4)\phi\,\partial^2\partial^2\phi\,,
\eeq
that vanish upon using the equations of motion \eqref{eom.4phi}. The corresponding scale and special conformal currents can be read through \eqref{currents.scale.special}.

In order to obtain $\Theta^{\m\n}$ one would have had to couple the action \eqref{dsnksdw} to Einstein gravity in a Weyl invariant way. This is equivalent to having to obtain the Weyl extension of the square of the Laplacian $\nabla^2$ (see e.g. \cite{Erdmenger:1997wy}). This procedure, however, does not provide a geometric interpretation for the improvement terms. We show below that coupling \eqref{dsnksdw} to MAG reveals the geometric origin of the improvements terms, and in fact it is rather straightforward. 

As before the main idea is to couple \eqref{dsnksdw} to MAG in a Weyl invariant manner. For that we need to construct  MAG Weyl-covariant derivatives on scalars $\Phi$ and a mixed tensors $X_{\mu_1\dots\mu_p}{}^{\nu_1\dots\nu_q}$ of corresponding weights $w$. We do that with the helps of the torsion vector $S_\m$ as
\beq
\begin{split}
&D_\mu \Phi=\left(\partial_\mu-\frac{2w}{d-1}S_\mu\right)\Phi\,,\\
&D_\mu X_{\mu_1\dots\mu_p}{}^{\nu_1\dots\nu_q}=\nabla_\mu X_{\mu_1\dots\mu_p}{}^{\nu_1\dots\nu_q}
-2\frac{w+p-q}{d-1}S_\mu X_{\mu_1\dots\mu_p}{}^{\nu_1\dots\nu_q}\,.
\end{split}
\eeq
These covariant derivatives  transforms under the following Weyl transformations
\beq
\label{skqseqs1}
g_{\mu\nu}\mapsto e^{-2\Omega}g_{\mu\nu}\,,\quad
\Phi\mapsto e^{w\Omega}\Phi\,,\quad X_{\mu_1\dots\mu_p}{}^{\nu_1\dots\nu_q}\mapsto e^{w\Omega}X_{\mu_1\dots\mu_p}{}^{\nu_1\dots\nu_q}\,,\quad
w=\frac{d-4}{2}\,,
\eeq
and the frame transformations of $S_\m$ in \eqref{frame.torsion.vector}
as
\beq
D_\mu \Phi\mapsto e^{w\Omega} D_\mu \Phi\,,\quad D_\mu X_{\mu_1\dots\mu_p}{}^{\nu_1\dots\nu_q}\mapsto e^{w\Omega} D_\mu X_{\mu_1\dots\mu_p}{}^{\nu_1\dots\nu_q}\,.
\eeq
Then, the following  action is manifestly Weyl and frame rescaling invariant.
\beq
\label{fkwskefdjw}
S_4=-\frac12\int d^d x\sqrt{-g}\, (D^2\phi)^2 \,,\quad D^2\phi=g^{\mu\nu}D_\mu D_\nu\phi\,,
\eeq
where
\beq
\label{fkwskefdjw1}
D^2\phi=\nabla^2\phi-\frac{2w}{d-1}\nabla_\lambda S^\lambda\phi-\frac{4w+2}{d-1}S^\lambda\partial_\lambda\phi
+\frac{4w(w+1)}{(d-1)^2}S_\lambda S^\lambda\phi\,.
\eeq
From the above, using \eqref{hyper.def} and \eqref{skqseqs1} we find that the hypermomentum tensor in the flat limit equals to
\beq
\label{sksns21}
\begin{split}
\Delta^{\nu\mu\lambda}&=-2\eta^{\mu\lambda}\partial^2\phi\partial^\nu\phi+\frac{d-4}{d-1}\left(\partial^\mu(\phi\partial^2\phi )\eta^{\lambda\nu}-\partial^\lambda(\phi\partial^2\phi )\eta^{\mu\nu}\right)\\
&-2\frac{d-3}{d-1}\partial^2\phi\left(\eta^{\lambda\nu}\partial^\mu\phi-\eta^{\mu\nu}\partial^\lambda\phi\right)\,,
\end{split}
\eeq
which of course differs from (\ref{sksns1}). 
Inserting \eqref{sksllskaq} and \eqref{sksns21} into \eqref{EMTs.flat}, we find
\beq
\label{hhyhydghh}
\begin{split}
t_{\mu\nu}&=2\partial_\mu\partial_\nu\phi\partial^2\phi-\frac12\eta_{\mu\nu}\,(\partial^2\phi)^2-\partial_\mu\left(\partial^2\phi\partial_\nu\phi\right)\\
&-\partial_\nu\left(\partial^2\phi\partial_\mu\phi\right)+\eta_{\mu\nu}\partial_\lambda(\partial^2\phi\partial^\lambda\phi)\\
&+\frac12\frac{d-4}{d-1}\left(\partial_\mu\partial_\nu-\eta_{\mu\nu}\partial^2\right)(\phi\partial^2\phi)\\
&+\frac{2}{d-1}(\partial_\nu(\partial_\mu\phi\partial^2\phi)-
\eta_{\mu\nu}\partial_\lambda(\partial^\lambda\phi\partial^2\phi))\,,
\end{split}
\eeq
where the terms in the second to fourth line are identically conserved line-by-line. In addition, starting from \eqref{hhyhydghh} we find
\beq
\label{canonical.scale.improved.prop}
\partial^\mu t_{\mu\nu}=-\partial_\nu\phi\,\partial^2\partial^2\phi\,,\quad
\eta^{\mu\nu}t_{\mu\nu}=-\frac{1}{2}(d-4)\phi\,\partial^2\partial^2\phi\,,
\eeq
hence upon using \eqref{eom.4phi} we find that $ t_{\mu\nu}$ is on-shell divergenceless and traceless. Comparing \eqref{hhyhydghh} with \eqref{Thetancs} we find that their difference corresponds to an off-shell improvement term regarding the conservation but also the trace. Thus, the two stress tensors are equivalent, yielding the same Noether charges up to boundary terms.

Although the action \eqref{dsnksdw} is Lorentz invariant the stress tensor \eqref{hhyhydghh} is non-symmetric 
\beq
t_{\mu\nu}-t_{\nu\mu}+\frac{4}{d-1}\partial_\lambda\left(\delta^\lambda{}_{[\mu}\partial_{\nu]}\phi\partial^2\phi\right)=0
\eeq
and using \eqref{ckdfkjdk} we read 
\beq
\label{hfksjsoaq}
\Sigma_{\mu\nu}{}^\lambda=\frac{4}{d-1}\delta^\lambda_{[\mu}\partial_{\nu]}\phi\partial^2\phi\,,
\eeq
whose trace is given by
\beq
\Sigma_{\lambda\nu}{}^\lambda=2\partial_\nu\phi\partial^2\phi\,,
\eeq
that satisfies
\beq
\label{BR.scale}
\Sigma_{\lambda\nu}{}^\lambda=\partial^\mu  Z_{\mu\nu}\,,\quad 
Z_{\mu\nu}=2\partial_\mu\phi\partial_\nu\phi-\eta_{\mu\nu}(\partial\phi)^2\,.
\eeq
Then, using \eqref{hfksjsoaq} we find
\beq
\Sigma_{\nu\lambda\mu}+\Sigma_{\mu\nu\lambda}+\Sigma_{\mu\lambda\nu}
=\frac{4}{d-1}\left(\eta_{\mu\nu}\partial_\lambda\phi\partial^2\phi-\eta_{\lambda\nu}\partial_\mu\phi\partial^2\phi\right)\,,
\eeq
so the corresponding Belinfante--Rosenfeld scale (BRS) energy--momentum tensor \eqref{BR1}, using \eqref{PsiSigma}, reads
\beq
\label{qpds12tu1}
\begin{split}
T^\text{BRS}_{\mu\nu}&=2\partial_\mu\partial_\nu\phi\partial^2\phi-\frac12\eta_{\mu\nu}\,(\partial^2\phi)^2-\partial_\mu\left(\partial^2\phi\partial_\nu\phi\right)\\
&-\partial_\nu\left(\partial^2\phi\partial_\mu\phi\right)+\eta_{\mu\nu}\partial_\lambda(\partial^2\phi\partial^\lambda\phi)\\
&+\frac12\frac{d-4}{d-1}\left(\partial_\mu\partial_\nu-\eta_{\mu\nu}\partial^2\right)(\phi\partial^2\phi)\\
&=T^\text{BR}_{\mu\nu}+\frac12\frac{d-4}{d-1}\left(\partial_\mu\partial_\nu-\eta_{\mu\nu}\partial^2\right)(\phi\partial^2\phi)\,,
\end{split}
\eeq
where in the last step we have employed \eqref{qpds12tu}. The tensor \eqref{qpds12tu1} is on-shell conserved, symmetric and its trace assumes the form 
\beq
\eta^{\mu\nu}T^\text{BRS}_{\mu\nu}=\eta^{\mu\nu}(t_{\mu\nu}+\partial_\lambda \Sigma_\mu{}^\lambda{}_\nu)
=-\frac{1}{2}(d-4)\phi\,\partial^2\partial^2\phi+\partial^\mu\partial^\nu  Z_{\mu\nu}\,,
\eeq
where we have used the second of \eqref{canonical.scale.improved.prop} and \eqref{BR.scale}. The latter formula ensures the on-shell conformal invariance of \eqref{dsnksdw} as we can construct the corresponding conserved, symmetric and traceless tensor $\Theta^{\mu\nu}$ using \eqref{sconf5}, where 
$(T^\text{BR}_{\mu\nu},\Lambda_{\mu\nu})\mapsto(T^\text{BRS}_{\mu\nu},-Z_{\mu\nu})$. The corresponding scale and special conformal currents can be read through \eqref{currents.scale.special}.

\subsection*{Six-derivative case}

Let us now consider the six-derivative free field also considered e.g. in~\cite{Osborn:2016bev} with action given by
\beq
\label{action.6derivative}
S_6=-\frac12\int d^d x\, (\partial_\mu\partial^2\phi)^2\,,\quad \partial^2\phi=\eta^{\mu\nu}\partial_\mu\partial_\nu\phi
\eeq
and the equations of motion are given by
\beq
\label{eom.6phi}
\partial^2\partial^2\partial^2\phi=0\,.
\eeq
Generalizing it to curved spacetimes we consider for metric connections the action
\beq
\label{action6.Riemann}
S_6=-\frac12\int d^d x\sqrt{-g}\, (\nabla_\mu\nabla^2\phi)^2\,,\quad \nabla^2\phi=\eta^{\mu\nu}\nabla_\mu\nabla_\nu\phi\,,
\eeq
leading to the metric stress tensor \eqref{grav.stress} and the hypermomentum tensor \eqref{hyper.def}
\beq
\label{metric.hyper.6der}
\begin{split}
&T_{\mu\nu}=\nabla_\mu\nabla^2\phi\nabla_\nu\nabla^2\phi-2\nabla_\mu\nabla_\nu\phi\nabla^2\nabla^2\phi-\frac12g_{\mu\nu}(\nabla_\lambda\nabla^2\phi)^2\,,\\
&\Delta_\nu{}^{\mu\lambda}=2g^{\lambda\mu}\nabla_\nu\phi\nabla^2\nabla^2\phi\,,
\end{split}
\eeq
where $\phi$ is a scalar field, that is $\nabla_\mu\phi=\partial_\mu\phi$. Using \eqref{EMTs.flat} for the trivial vacuum we easily find the non-symmetric stress tensor
\beq
\label{action.6derivative.can}
t_{\mu\nu}=\partial_\mu\partial^2\phi\partial_\nu\partial^2\phi-2\partial_\mu\partial_\nu\phi\partial^2\partial^2\phi-\frac12g_{\mu\nu}(\partial_\lambda\partial^2\phi)^2+\partial_\mu(\partial_\nu\phi\partial^2\partial^2\phi)\,,
\eeq
that coincides with the canonical stress tensor for the action \eqref{action.6derivative}. The tensor \eqref{action.6derivative.can} is on-shell conserved since
\beq
\partial^\mu t_{\mu\nu}=\partial_\nu\phi\partial^2\partial^2\partial^2\phi\,.
\eeq
Using the Belinfante--Rosenfeld procedure \eqref{BR1} or considering the action \eqref{action6.Riemann} for a Levi--Civita connection we find
\beq
\label{Belinfante.6der}
\begin{split}
T^\text{BR}_{\mu\nu}&=\partial_\mu\partial^2\phi\partial_\nu\partial^2\phi-2\partial_\mu\partial_\nu\phi\partial^2\partial^2\phi-\frac12g_{\mu\nu}(\partial_\lambda\partial^2\phi)^2+\partial_\mu(\partial_\nu\phi\partial^2\partial^2\phi)\\
&+\partial_\nu(\partial_\mu\phi\partial^2\partial^2\phi)-g_{\mu\nu}\partial_\lambda(\partial^\lambda\phi\partial^2\partial^2\phi)\,,
\end{split}
\eeq
where the second line is identically conserved.

The action \eqref{action6.Riemann} is invariant under the global transformations
\beq
g_{\mu\nu}\mapsto e^{-2\Omega}g_{\mu\nu}\,,\quad
\phi\mapsto e^{w\Omega}\phi\,,\quad w=\frac{d-6}{2}\,.
\eeq
As in the four-derivative case we make the symmetry local using as a Weyl connection the torsion vector reaching the expression for the hypermomentum tensor
\beq
\begin{split}
\Delta_\nu{}^{\mu\lambda}&=2\eta^{\lambda\mu}\partial_\nu\phi\partial^2\partial^2\phi-
\frac{d-6}{d-1}\left(\delta_\nu^\lambda\partial^\mu-\delta_\nu^\mu\partial^\lambda\right)\left(\phi\partial^2\partial^2\phi\right)\\
&+2\frac{d-5}{d-1}\left(\delta_\nu^\lambda\partial^\mu\phi-\delta_\nu^\mu\partial^\lambda\phi\right)\partial^2\partial^2\phi
-\frac{d-2}{d-1}\left(\delta_\nu^\lambda\partial^\mu\partial^2\phi-\delta_\nu^\mu\partial^\lambda\partial^2\phi\right)\partial^2\phi\,.
\end{split}
\eeq
Using the latter and the metric tensor in \eqref{metric.hyper.6der} we find scale canonical stress tensor
\beq
\begin{split}
t_{\mu\nu}&=\partial_\mu\partial^2\phi\partial_\nu\partial^2\phi-2\partial_\mu\partial_\nu\phi\partial^2\partial^2\phi-\frac12g_{\mu\nu}(\partial_\lambda\partial^2\phi)^2+\partial_\mu(\partial_\nu\phi\partial^2\partial^2\phi)\\
&-\frac12\frac{d-6}{d-1}\left(\partial^\mu\partial^\nu-\eta^{\mu\nu}\partial^2\right)\left(\phi\partial^2\partial^2\phi\right)\\
&+\frac{d-5}{d-1}\left(\partial_\nu(\partial_\mu\phi\partial^2\partial^2\phi)-\eta_{\mu\nu}\partial_\lambda(\partial^\lambda\phi\partial^2\partial^2\phi)\right)\\
&-\frac12\frac{d-2}{d-1}\left(\partial_\nu(\partial_\mu\partial^2\phi\partial^2\phi)-\eta_{\mu\nu}\partial_\lambda(\partial^\lambda\partial^2\phi\partial^2\phi)\right)\,,
\end{split}
\eeq
satisfying
\beq
\label{canonical.scale.improved.prop1}
\partial^\mu t_{\mu\nu}=\partial_\nu\phi\partial^2\partial^2\partial^2\phi\,,\quad \eta^{\mu\nu}t_{\mu\nu}=\frac12(d-6)\phi\partial^2\partial^2\partial^2\phi\,.
\eeq 
The latter stress tensor is not a symmetric tensor satisfying \eqref{ckdfkjdk}, where 
\beq
\Sigma_{\mu\nu\lambda}=-\frac{8}{d-1}\eta_{\lambda[\mu}\partial_{\nu]}\phi\partial^2\partial^2\phi
-\frac{d-2}{d-1}\eta_{\lambda[\mu}\partial_{\nu]}\partial^2\phi\partial^2\phi\,.
\eeq
Its trace reads
\beq
\label{BR.scale1}
\Sigma_{\lambda\nu}{}^\lambda=-4\partial_\nu\phi\partial^2\partial^2\phi-\frac12(d-2)\partial_\nu\partial^2\phi\partial^2\phi=\partial^\mu Z_{\mu\nu}\,,
\eeq
where (see also \cite{Stergiou:2022qqj})
\beq
Z_{\mu\nu}=2\left(\phi\partial_\mu\partial_\nu\partial^2\phi+\partial_\mu\partial_\nu\phi\partial^2\phi\right)
-2\left(\eta_{\mu\nu}\phi\partial^2\partial^2\phi+2\partial_{(\mu}\phi\partial_{\nu)}\partial^2\phi\right)-\frac12(d-2)\eta_{\mu\nu}\left(\partial^2\phi\right)^2\,.
\eeq
We can now define the BRS stress tensor \eqref{BR1}, using \eqref{PsiSigma}, reads
\beq
\label{Belinfante.6der.scale}
T_{\mu\nu}^\text{BRS}=T_{\mu\nu}^\text{BR}-\frac12\frac{d-6}{d-1}\left(\partial_\mu\partial_\nu-\eta_{\mu\nu}\partial^2\right)\left(\phi\partial^2\partial^2\phi\right)\,,
\eeq
where $T_{\mu\nu}^\text{BR}$ was given in \eqref{Belinfante.6der}. The trace of reads
\beq
\eta^{\mu\nu}T^\text{BRS}_{\mu\nu}=\eta^{\mu\nu}(t_{\mu\nu}+\partial_\lambda \Sigma_\mu{}^\lambda{}_\nu)
=\frac{1}{2}(d-6)\phi\,\partial^2\partial^2\partial^2\phi+\partial^\mu\partial^\nu  Z_{\mu\nu}\,,
\eeq
where we have used the second of \eqref{canonical.scale.improved.prop1} and \eqref{BR.scale1}. The latter formula ensures the on-shell conformal invariance of \eqref{action.6derivative} as we can construct the corresponding conserved, symmetric and traceless tensor $\Theta^{\mu\nu}$ using \eqref{sconf5}, where 
$(T^\text{BR}_{\mu\nu},\Lambda_{\mu\nu})\mapsto(T^\text{BRS}_{\mu\nu},-Z_{\mu\nu})$. The corresponding scale and special conformal currents can be read through \eqref{currents.scale.special}.

Similarly we can extend the above treatment for other higher-derivative free fields
\beq
S_{4n}=-\frac{1}{2}\int d^dx(\partial^{2n}\phi)^2\,,\quad
S_{4n+2}=-\frac{1}{2}\int d^dx(\partial_\mu\partial^{2n}\phi)^2\,,\quad n\geqslant1\,.
\eeq

\section{Conclusion and Outlook}
\label{Conclusion.Outlook}

The Belinfante--Rosenfeld and conformal improvements of the canonical e.m. tensor of matter theories in flat spacetimes are complicated but straightforward procedures. In most cases they are circumvented by coupling the theory to Einstein or Weyl gravity, but this leaves open the question regarding the geometrical origin of the process. In this work we have provided a geometric origin of the improvement processes by extending the coupling of the matter system to metric-affine geometry, namely allowing for connection-matter couplings. We have  explicitly shown in a number of examples that the improvement terms are directly related to the hypermomentum of the matter theory. Nevertheless, this relationship is theory dependent and in the case of conformal improvement terms there does not appear to be a general formula although we have demonstrated it in some non trivial cases. 

Our main result is that by coupling the matter theory to MAG rather than just Einstein gravity and then taking the flat limit we arrive at a geometric interpretation of the improvement terms in flat space, both for Lorentz as well as for conformal invariance, in terms of the hypermomentum. Moreover, the coupling of the matter theory to MAG can be done in a minimal way and this avoids the complicated procedure of Weyl extending differential operators in Einstein gravity. The point is that to obtain the flat limit of the hypermomentum we just need to know the linear terms in the affine connection in any MAG extension of the matter theory. This is a significant simplification; however, it is not without ambiguity in the case of a MAG Weyl extension. In this work, we have considered the simple case of local frame rescaling, and we have shown in explicit nontrivial examples how the hypermomentum determines the improvement term. 

Extensions of our work may offer a new perspective on the geometric and gravitational description of matter systems invariant under subsets or limits of Poincare or conformal symmetries. Such systems may arise in AdS/CFT holography or in fluid dynamics. In particular, since coupling to MAG gives us access to the canonical e.m. tensor of the matter theory, we  can describe matter systems with less symmetries e.g. systems without Lorentz symmetry, or systems with scalar but not conformal symmetry. Looking at our main equations \eqref{ksi} and \eqref{MAGtrace} we note that had we considered non-trivial flat limits i.e. situations where the  connection does not vanish while the metric stays flat, then we could describe matter systems whose canonical e.m. tensor cannot be improved to be symmetric or traceless.\footnote{The possibility that Lorentz symmetry breaking terms might have geometric origins was discussed also in \cite{Kiritsis:2012ta}.} Other possibilities include the description of Carrollian, Galilean or even Aristotelian matter theories, e.g. \cite{deBoer:2017ing,Ciambelli:2018xat,deBoer:2020xlc,Petkou:2022bmz,deBoer:2023fnj}. We plan to return in these questions in the near future.

\section*{Acknowledgements}

We would like to thank L. Ciambelli, A. Fiorucci, F. Hehl, S. Pekar, M. Petropoulos, A. Stergiou and M. Vilatte for useful discussions and correspondence. The work of DI is supported by the Estonian Research Council via the Center of Excellence `\emph{Foundations of the Universe}' (TK202U4). The work of MK is supported by the U.S. Department of Energy, Office of Science, Office of High Energy Physics under award number DESC0015655. ACP and KS. would like to thank the Erwin Schr\"odinger International Institute for Mathematics and Physics for hospitality and financial support during the "Carrollian Physics and Holography" workshop, $2^\text{nd}$ to $26^\text{th}$ April 2024, Vienna. Preliminary results of the present work were presented in a talk by KS, at that workshop. KS would like also to thank the Institute of Physics of the University of Humboldt for hospitality and financial support, where results of the present work were also presented.

\appendix

\section{Metric-affine gravity}
\label{Appendix.MAG}

We shall work in the metric-affine framework. To this end we consider a d-dimensional manifold of Lorentzian signature which we endow with the line element $ds^2=g_{\mu\nu}dx^{\mu}dx^{\nu}$, $\mu,\nu=0,1,2,\ldots,d-1$ and a generic affine connection $\nabla$ with components $\Gamma^{\lambda}{}_{\mu\nu}$.  The covariant derivative acts on a mixed (1,1) tensor, say $A^\mu{}_\nu$ as 
\beq
\label{fjeoqq}
\nabla_{\kappa}A^\mu{}_\nu=\partial_{\kappa}A^\mu{}_\nu+\Gamma^{\mu}{}_{\lambda\kappa}
A^\lambda{}_\nu-
\Gamma^{\lambda}{}_{\nu\kappa}
A^\mu{}_\lambda\,.
\eeq
We also define the torsion and curvature of the connection as usual through
\beq
[\nabla_\kappa,\nabla_\lambda]A^\mu{}_\nu=
R^\mu{}_{\rho\kappa\lambda}A^\rho{}_\nu-
R^\rho{}_{\mu\kappa\lambda}A^\mu{}_\rho+2S_{\kappa\lambda}{}^\rho\nabla_\rho A^\mu{}_\nu\,,
\eeq
where
\beq
\label{torsion1}
S_{\mu\nu}{}{}^{\lambda}:=\Gamma^{\lambda}{}{}_{[\mu\nu]}\,,\quad 
S_{\mu\nu\rho}=g_{\rho\lambda}S_{\mu\nu}{}^\lambda
\eeq
and 
\begin{equation}
R^{\mu}{}_{\nu\alpha\beta}:= 2\partial_{[\alpha}\Gamma^{\mu}{}_{|\nu|\beta]}+2\Gamma^{\mu}{}_{\rho[\alpha}\Gamma^{\rho}{}_{|\nu|\beta]}\,,
\end{equation}
respectively and $\nu$ is excluded from antisymmetry on curvature tensor. From the curvature tensor we can define three kinds of contractions, namely
\beq
R_{\nu\beta}=R^{\mu}{}_{\nu\mu\beta}\,,\quad
\hat R_{\alpha\beta}=R^{\mu}{}_{\mu\alpha\beta}\,,\quad
\check R^{\mu}{}_\beta=
R^{\mu}{}_{\nu\alpha\beta}g^{\nu\alpha}
\eeq
and their corresponding contractions
\beq
R=g^{\mu\nu}R_{\mu\nu}=-g^{\mu\nu}\check R_{\mu\nu}\,,\quad 
g^{\mu\nu}\hat R_{\mu\nu}=0\,.
\eeq
Obviously these objects need only a connection to be defined and do not require a metric. On the other hand a metric (along with the connection) is essential to construct the non-metricity tensor which we define via,
\beq
\label{jxcsjfhsd}
Q_{\rho\mu\nu}=-\nabla_{\rho}g_{\mu\nu}=-\partial_{\rho}g_{\mu\nu}+\Gamma^{\lambda}{}_{\mu\rho}\,g_{\lambda\nu}+\Gamma^{\lambda}{}_{\nu\rho}\,g_{\mu\lambda}\,,
\eeq
that measures the failure of the metric to be covariantly constant. Out of non-metricity we can construct two vectors by the following contractions
\beq
Q_{\mu}:=Q_{\mu\alpha\beta}g^{\alpha\beta}, \quad q_{\mu}:=Q_{\alpha\nu\mu}g^{\alpha\nu}.
\eeq
As for torsion we can define a vector as follows\footnote{Along with a pseudo-vector $t_{\mu}:=\epsilon_{\mu\nu\alpha\beta}S^{\nu\alpha\beta}$, defined only for $n=4$.}
\beq
\label{torsion2}
S_{\mu}:=S_{\mu\alpha}{}^{\alpha}.
\eeq
 Note that with the above definitions of torsion and non-metricity the affine connection can be decomposed as
 (see for instance, \cite{Iosifidis:2019dua}) 
\begin{equation}\label{decgamma}
{\Gamma^\lambda}_{\mu \nu} = \tilde{\Gamma}^\lambda_{\phantom{\lambda} \mu \nu} + {N^\lambda}_{\mu \nu}\,,
\end{equation}
where 
\beq\label{lcconn}
\tilde{\Gamma}^\lambda_{\phantom{\lambda}\mu\nu} = \frac12 g^{\rho\lambda}\left(\partial_\mu 
g_{\nu\rho} + \partial_\nu g_{\rho\mu} - \partial_\rho g_{\mu\nu}\right)\,,
\eeq
is the usual Levi--Civita connection constructed from $g_{\mu\nu}$ and the additional tensor ${N^\lambda}_{\mu\nu}$ is defined through
\beq\label{distortion}
\begin{split}
{N^\lambda}_{\mu\nu} &= {\frac12 g^{\rho\lambda}\left(Q_{\mu\nu\rho} + Q_{\nu\rho\mu}
- Q_{\rho\mu\nu}\right)} - {g^{\rho\lambda}\left(S_{\rho\mu\nu} +
S_{\rho\nu\mu} - S_{\mu\nu\rho}\right)} \,.
\end{split}
\eeq	
 is the so-called distortion tensor, measuring the departure from the Riemannian geometry.
 Out of the distortion we can extract three traces as $N^{(1)}_{\nu}:=N^{\alpha}{}_{\alpha\nu}$, $N^{(2)}_{\nu}:=N^{\alpha}{}_{\nu\alpha}$ and $N^{(3)}_{\nu}:=N_{\alpha\mu\nu}g^{\mu\nu}$ which are linear combinations of the torsion and non-metricity vectors as can be trivially verified.

\section{Canonical energy--momentum tensor}
\label{Appendix.Canonical}

Let us now prove here the known fact that the canonical energy--momentum tensor defined by the variation with respect to the vielbein as it is read from Eq.\eqref{cemt} 
\beq
t^{\mu}{}_{\nu}=t^{\mu}{}_{a}e_{\nu}{}^{a}
=\frac{1}{\sqrt{-g}}\frac{\delta (\sqrt{-g}\mathcal{L}_{M})}{\delta e_{\mu}{}^{a}}e_{\nu}{}^{a} \label{tcan}
\eeq
coincides the tensorial  canonical expression \eqref{tauordinary}.

We first employ the chain rule on (\ref{tcan}) we find
\begin{gather}
t^{\mu}{}_{\nu}=\frac{1}{\sqrt{-g}}\frac{\partial (\sqrt{-g}\mathcal{L}_{M})}{\partial e^{\lambda}{}_{c}}\frac{\partial e^{\lambda}{}_{c}}{\partial e_{\mu}{}^{a}}e_{\nu}{}^{a}=-\frac{1}{\sqrt{-g}}\delta_{\nu}^{\lambda}e^{\mu}{}_{c}\frac{\partial (\sqrt{-g}\mathcal{L}_{M})}{\partial e^{\lambda}{}_{c}}= \nonumber \\
=-e^{\mu}{}_{c}\left\{\frac{1}{\sqrt{-g}}\frac{\partial \sqrt{-g}}{\partial e^{\nu}{}_{c}}\mathcal{L}_{M}+\frac{\partial \mathcal{L}_{M}}{\partial e^{\nu}{}_{c}  }  \right\}\,,
\label{tdU.dkskls}
\end{gather}
where in the second equality we used the identity
\beq
\frac{\partial e^{\lambda}{}_{c}}{\partial e_{\mu}{}^{a}}=-e^{\mu}{}_{c}e^{\lambda}{}_{a}\,.
\eeq
Then, from the  relation $g_{\mu\nu}=e_{\mu}{}^{a}e_{\nu}{}^{b}\eta_{ab}$ we easily obtain 
\beq
\frac{\partial \sqrt{-g}}{\partial e^{\nu}{}_{c}}=-\sqrt{-g}e_{\nu}{}^{c}
\eeq
and this to \eqref{tdU.dkskls} it follows that
\beq
t^{\mu}{}_{\nu}=\delta^{\mu}_{\nu}\mathcal{L}_{M}-e^{\mu}{}_{c}\frac{\partial \mathcal{L}_{M}}{\partial e^{\nu}{}_{c}  } 
\eeq
We now need only to take care of the last term on the right-hand  side of the latter. Since the vielbein appears only through the covariant differentiation of the matter fields, we have
\beq
\nabla_{a}\phi^A=e^{\rho}{}_{a}\nabla_{\rho}\phi^A
\eeq
and therefore
\beq
\frac{\partial (\nabla_{a}\phi^A)}{\partial e^{\lambda}{}_{c}}=\delta^{c}_{a}\nabla_{\lambda}\phi^A 
\eeq
then on the premise that the matter action depends algebraically on the vielbein (i.e. not on its derivatives), employing the chain rule again and using \eqref{derL}, we find
\beq
\frac{\delta \mathcal{L}_{M}}{\delta e^{\nu}{}_{c}  }=\frac{\partial\mathcal{L}_{M}}{\partial(\nabla_{a}\phi^A) }\frac{\partial(\nabla_{a}\phi^A)}{\partial e^{\nu}{}_{c}}=\frac{\partial\mathcal{L}_{M}}{\partial (\nabla_{c}\phi^A)}\nabla_{\nu}\phi^A
\eeq
Upon inserting the above into \eqref{tL} we  establish the claimed equivalence of \eqref{tcan} with \eqref{tauordinary}
\beq
t^{\mu}{}_{\nu}
=-e^{\mu}{}_{c}\frac{\partial\mathcal{L}_{M}}{\partial \left(\nabla_{c}\phi^A\right) }\nabla_{\nu}\phi^A+\delta^{\mu}_{\nu}\mathcal{L}_{M}
=-\frac{\partial \mathcal{L}_{M}}{\partial (\nabla_{\mu}\phi^A)}\nabla_{\nu}\phi^A+\delta^{\mu}_{\nu}\mathcal{L}_{M}\equiv t_c^{\mu}{}_{\nu}\,.
\eeq

\section{Proof of the equation \eqref{EMTs}}
\label{Appendix.proof}

Let us start by \eqref{cemt} and use the chain rule
\beq
\label{vdsdnsw0}
{t^a}_\mu(x) := -\frac{1}{\sqrt{-g}} \frac{\delta S_{\text{M}}}{\delta {e^\mu{}_a}(x)}=
-\frac{1}{\sqrt{-g}} \int d^d y\left\{\frac{\delta S_{\text{M}}}{\delta g^{\kappa\lambda}(y)}\frac{\delta g^{\kappa\lambda}(y)}{\delta {e^\mu{}_a}(x)}+
\frac{\delta S_{\text{M}}}{\delta \Gamma^\rho{}_{\kappa\lambda}(y)}\frac{\delta \Gamma^\rho{}_{\kappa\lambda}(y)}{\delta {e^\mu{}_a}(x)}\right\}\,.
\eeq
To proceed we employ $
g^{\kappa\lambda}=\eta^{ab}e^\kappa{}_a e^\lambda{}_b
$ and \eqref{spin.constraint} \footnote{Assuming that the spin-connection $\omega_{ab\mu}$ is independent of $e^\mu{}_a$ 
 and using the identity $x\,\delta'(x)=-\delta(x)$.},  yielding the expression for $t^\nu{}_\mu={t^a}_\mu e^\nu{}_a$ 
\beq
t^\nu{}_\mu=T^\nu{}_\mu
+\frac12\Delta_\mu{}^{\kappa\lambda}\Gamma^\nu{}_{\kappa\lambda}-\frac12\Delta_\rho{}^{\nu\lambda}\Gamma^\rho{}_{\mu\lambda}-\frac12\Delta_\mu{}^{\kappa\lambda}e^\nu{}_a\partial_\lambda e^a{}_\kappa
+\frac{1}{2\sqrt{-g}}\partial_\rho\left(\sqrt{-g}\Delta_\mu{}^{\kappa\rho}e^a{}_\kappa\right)e^\nu{}_\alpha\,.
\eeq
where we have used the definitions \eqref{grav.stress} and \eqref{hyper.def}. To proceed we use \eqref{fjeoqq} for $\Delta_\mu{}^{\nu\rho}$
yielding after some algebra
\beq
\label{nfkdss0}
t^\nu{}_\mu=T^\nu{}_\mu+
\frac12\nabla_\lambda\Delta_\mu{}^{\nu\lambda}
+\frac12\left(\partial_\rho\ln\sqrt{-g}-\Gamma^\lambda{}_{\rho\lambda}\right)\Delta_\mu{}^{\nu\rho}\,.
\eeq
Then, using \eqref{torsion1} an \eqref{jxcsjfhsd}, we obtain
\beq
\label{nfkdss}
\Gamma^\lambda{}_{\rho\lambda}=\frac12g^{\lambda\sigma}\partial_\rho g_{\lambda\sigma}+\frac12Q_\rho+2S_\rho\quad\Longrightarrow\quad
\partial_\rho\ln\sqrt{-g}-\Gamma^\lambda{}_{\rho\lambda}=
\nabla_\rho\ln\sqrt{-g}-2S_\rho\,,
\eeq
where 
\beq
\label{nabla.density}
\frac12g^{\lambda\sigma}\partial_\rho g_{\lambda\sigma}=
\partial_\rho\ln\sqrt{-g}\,,\quad Q_\rho=-2\nabla_\rho\ln\sqrt{-g}\,,
\eeq
that can be easily found through \eqref{jxcsjfhsd}, \eqref{lcconn}. Employing \eqref{nfkdss} into \eqref{nfkdss0} we find
\beq
\label{vdsdnsw}
t^{\nu}{}_{\mu}
	= T^{\nu}{}_{\mu}+\frac{1}{2 \sqrt{-g}}\hat{\nabla}_\lambda(\sqrt{-g}\Delta_{\mu}{}^{\nu\lambda})\,,
\eeq
where we recall $\hat{\nabla}_{\lambda}=\nabla_{\lambda}-2 S_{\lambda}$, which is in agreement with \eqref{EMTs}.

\section{Revisiting known examples}
\label{Appendix.Examples}

In this Appendix we revisit the derivation of the canonical   energy--momentum tensor and the corresponding improvement terms in various known examples.

\subsection{Conformally coupled scalar}
\label{Conf.Appendix}

Let us consider the action of the conformally coupled scalar~\cite{Callar.Jackiw}, given in \eqref{action.conf} and restated here for reader's convenience
\beq
S_\text{cs}=\int d^dx\sqrt{-g}\left(-\frac12(\partial\phi)^2-\frac18\frac{d-2}{d-1}\ \tilde R\phi^2\right)\,,
\eeq
where $(\partial\phi)^2=g^{\mu\nu}\partial_\mu\phi\partial_\nu\phi$ and $\tilde R$ is the Ricci-scalar of the metric $g_{\mu\nu}$. The action is
invariant under the Weyl transformation \eqref{skqseqs}
\beq
\label{Weyl.scalar}
g_{\mu\nu}\mapsto e^{-2\Omega} g_{\mu\nu}\,,\quad
\phi\mapsto e^{\frac{d-2}{2}\Omega}\phi\,,
\eeq
where the Ricci scalar transforms as \eqref{Weyl.Ricci}.
Working out the gravitational stress energy tensor \eqref{grav.stress} we find
\beq
\label{stress.conf.scalar}
T_{\mu\nu}=
\partial_\mu\phi\partial_\nu\phi-\frac12g_{\mu\nu}(\partial\phi)^2+
\frac14\frac{d-2}{d-1}\left(\tilde G_{\mu\nu}\phi^2+g_{\mu\nu}\tilde\nabla^2\phi^2-\tilde\nabla_\mu\tilde\nabla_\nu\phi^2\right)\,,
\eeq
where $\tilde\nabla$ is the covariant derivative constructed out of the Levi--Civita tensor for the metric $g_{\mu\nu}$ and $\tilde G_{\mu\nu}=\tilde R_{\mu\nu}-\frac12g_{\mu\nu}
\tilde R$ is the corresponding Einstein tensor.
This tensor is conserved and traceless upon using the equations of motion of \eqref{action.conf}
\beq
\Delta_2\phi=0\,,\quad \Delta_2=\tilde\nabla^2-\frac14\frac{d-2}{d-1}\tilde R\,,
\eeq
where $\Delta_2$ is the conformal Laplacian or the Yamabe operator. Under the Weyl transformation \eqref{Weyl.scalar} the Yamabe operator transforms as
\beq
\Delta_2\phi\mapsto e^{\frac{d+2}{2}\Omega}\Delta_2\phi\,,
\eeq
yielding the Weyl invariant action
\beq
S_\text{cs}=\frac12\int d^dx\sqrt{-g}\phi\Delta_2\phi\,,
\eeq
which up to a boundary term is equivalent to \eqref{action.conf}. Finally, upon taking the flat limit in \eqref{stress.conf.scalar} we readily find 
\beq
\label{stress.conf.scalar.flat}
T_{\mu\nu}=\partial_{\mu}\phi \partial_{\nu}\phi -\frac{1}{2}\eta_{\mu\nu}(\partial \phi)^{2} 
+\frac{1}{4}\frac{d-2}{d-1}\left(\eta_{\mu\nu} \partial^2\phi^2-\partial_\mu\partial_\nu\phi^2\right)\,,
\eeq
which coincides with \eqref{skkn}.

\subsection{The electromagnetic field}
\label{Maxwellapp}

Let us consider Maxwell field whose action is given by \eqref{Maxwell.action} and the energy--momentum tensor $t^\mu{}_\nu$ is given by
\eqref{canonicalEM} that is not symmetric, although the action \eqref{Maxwell.action}  is invariant under Lorentz transformation.
To symmetrize \eqref{canonicalEM} we follow BR procedure, (see \cite{Munoz:1996wp}) and we demand at first that the Maxwell action \eqref{Maxwell.action} is invariant under local Lorentz transformations \eqref{Lorentz.transf} under which
\beq
\delta_\omega A^\mu=-
\omega^\mu{}_\nu A^\nu\,.
\eeq
Doing so we find the spin-current \eqref{fksksaal} where
\beq
\label{qgeuwuw}
\Sigma^{\mu\nu\rho}=A^\rho F^{\mu\nu}-
A^\nu F^{\mu\rho}\,,
\eeq
obeying \eqref{hyper.Spin} with the hypermomentum given in \eqref{nxhkqo340}.
Starting from \eqref{BR1} and using \eqref{PsiSigma}, \eqref{canonicalEM}, \eqref{qgeuwuw} we find
\beq
\label{Belin.Maxwell}
\begin{split}
T_\text{BR}^{\mu\nu}&=t^{\mu\nu}-\partial_\lambda\left(A^\nu F^{\mu\lambda}\right)=F^{\mu\lambda}F^{\nu}{}_{\lambda}-\frac{1}{4}\eta^{\mu\nu}
F_{\kappa\lambda}F^{\kappa\lambda}\,,
\end{split}
\eeq
which coincides with \eqref{stressMaxwell}.

\noindent
Equivalently we could have considered the Maxwell field in a curved background, whose action is given by
\beq
\label{Maxwell.action.curved}
S_\text{EM}=\int 
d^dx\sqrt{-g}\mathcal{L}_\text{EM}\,,\quad
\mathcal{L}_\text{EM}=-\frac{1}{4}F_{\mu\nu}F^{\mu\nu}
\eeq
and upon using \eqref{grav.stress} and then restricting to flat we find agreement with \eqref{Belin.Maxwell}.

\subsection{Dirac field }
\label{Diracapp}

Let us consider the massive Dirac field in a $d$ dimensional spacetime described by the Hermitian action, flat limit of \eqref{Dirac.action.curv}
\beq
\label{Dirac.action}
S_\text{D}=\int d^dx {\cal L}_\text{D}\,,\quad
{\cal L}_\text{D}=-\frac{i}{2}(\bar\psi\gamma^\mu\partial_\mu\psi-\partial_\mu\bar\psi\gamma^\mu\psi)-im\bar\psi\psi\,,\quad \bar\psi=\psi^\dagger\gamma^0\,,
\eeq
whose equations of motion read
\beq
\label{osldsld}
(\gamma^\mu\partial_\mu+m)\psi=0\,,\quad 
\partial_\mu\bar\psi \gamma^\mu-m\bar\psi=0\,.
\eeq
Demanding invariance under a local translation
$
\delta_\varepsilon x^\mu=-\varepsilon^\mu(x)\,,
$
we easily find
\beq
\delta_\varepsilon S_\text{D}=-\int d^dx\,\partial_\mu\varepsilon_\nu t^{\mu\nu}\,,
\eeq
where the off-shell conserved canonical energy--momentum tensor is given by 
\beq
t^{\mu\nu}=
\frac{i}{2}\left(\bar\psi\gamma^\mu\partial^\nu\psi-\partial^\nu\bar\psi\gamma^\mu\psi\right)
+\eta^{\mu\nu}{\cal L}_D\,,
\eeq
also read through \eqref{tmnpi}.
Upon using the equations of motion \eqref{osldsld} it simplifies to 
\beq
\label{vdkdsksk}
t^{\mu\nu}=
\frac{i}{2}\left(\bar\psi\gamma^\mu\partial^\nu\psi-\partial^\nu\bar\psi\gamma^\mu\psi\right)\,.
\eeq
This is not symmetric despite the action \eqref{Dirac.action} been invariant under Lorentz transformations. To obtain a symmetric stress energy tensor we
follow BR procedure and we demand at first that the Dirac action  \eqref{Dirac.action} is invariant under local Lorentz transformations \eqref{Lorentz.transf} where the spinor transforms as \eqref{Lorentz.transf.spinor}. Demanding that the Dirac action \eqref{Dirac.action} is invariant under local Lorentz transformations \eqref{Lorentz.transf}
\beq
\delta_\alpha S_\text{D}=-\frac12\int d^dx\,
\partial_\mu\omega_{\nu\rho}(J^\mu)^{\nu\rho}\,,
\eeq
where the conserved spin-currents 
$\partial_\mu (J^\mu)^{\nu\rho}=0$,
are given by
\beq
(J^\mu)^{\nu\rho}=
t^{\mu\nu}x^\rho-t^{\mu\rho}x^\nu
-\Sigma^{\mu\nu\rho}\,,
\eeq
where
\beq
\label{hdksksks}
\Sigma^{\mu\nu\rho}=\frac{i}{2}\bar\psi\left(\gamma^\mu\Sigma^{\nu\rho}+\Sigma^{\nu\rho}\gamma^\mu\right)\psi\,,
\eeq
that coincides with the hypermomentum \eqref{Hyper.Dirac}
\beq
\Sigma^{\mu\nu\rho}=\Delta^{\rho\mu\nu}\,,
\eeq
trivially satisfying \eqref{hyper.Spin}, since it
is totally antisymmetric due to \eqref{Hyper.Dirac1}.

Lastly, it can be easily seen that the BR stress energy tensor $T_\text{BR}^{\mu\nu}$ defined in \eqref{BR1} coincides with the tensor in \eqref{improved.mostly.plus}. Indeed starting from \eqref{BR1}, \eqref{PsiSigma} and \eqref{EMTs.flat} we easily find 
\beq
\begin{split}
T_\text{BR}^{\mu\nu}&=t^{\mu\nu}+\frac12\partial_\lambda\Sigma^{\lambda\mu\nu}=\tau^{\mu\nu}-\frac12\partial_\lambda\Delta^{\nu\mu\lambda}=T^{\mu\nu}\\
&=\frac{i}{4}
\left(\bar\psi\gamma^\mu\partial^\nu\psi-
\partial^\nu\bar\psi\gamma^\mu\psi\right)
+
\frac{i}{4}
\left(\bar\psi\gamma^\nu\partial^\mu\psi-
\partial^\mu\bar\psi\gamma^\nu\psi\right)\,,
\end{split}
\label{BR.Dirac}
\eeq
where in the last step we have used \eqref{improved.mostly.plus}.
In passing we note that the derived BR stress energy tensor is given by the symmetrisation of the canonical one \eqref{vdkdsksk}, where the additional term is on-shell conserved.

\subsection{Non-unitary scalar CFTs}
\label{nonunitaryapp}

Let us consider the non-unitary scalar CFT (conformal field theory) described by the action \eqref{dsnksdw} and the canonical energy--momentum tensor $t_{\mu\nu}$ is given by
\eqref{Noether.general} that is not symmetric, despite the action been invariant under Lorentz transformations. To symmetrize \eqref{Noether.general} we follow BR procedure and we demand at first that the action  \eqref{dsnksdw} is invariant under local Lorentz transformations. Doing so we find the on-shell conserved spin-current \eqref{fksksaal} where
\beq
\label{dsnksdwww}
\Sigma^{\mu\nu\rho}=\partial^2\phi\left(\eta^{\mu\nu}\partial^\rho\phi-\eta^{\mu\rho}\partial^\nu\phi\right)\,,
\eeq
obeying \eqref{hyper.Spin}, where the hypermomentum was given in \eqref{sksns1}. Finally, starting from \eqref{BR1} and using \eqref{PsiSigma}, \eqref{dsnksdwww} we reach \eqref{qpds12tu}, which is restated here for reader's convenience
\beq
\label{qpds12tu0}
\begin{split}
T_\text{BR}^{\mu\nu}&=2\partial^\mu\partial^\nu\phi\partial^2\phi-\frac12\eta^{\mu\nu}\,\partial^2\phi \partial^2\phi-\partial^\mu\left(\partial^2\phi\partial^\nu\phi\right)\\
&-\partial^\nu(\partial^2\phi\partial^\mu\phi)+\eta^{\mu\nu}\partial_\lambda(\partial^2\phi\partial^\lambda\phi)\,.
\end{split}
\eeq

\section{Improved Noether current}
\label{Sec:Improved}

In this Appendix, we will consider improvements of the conserved currents where analogue considerations have been considered in
\cite{Gross:1970tb} and see also \cite{Osborn.CFTs,Brauner:2019lcb,Freese:2021jqs,Kourkoulou:2022ajr,Gieres:2022cpn,Kim:2024ewt}.

Let us consider the action in flat spacetime
\beq
S=\int d^dx{\cal L}(x)\,,
\eeq
and varying $x^\mu$ as 
\beq
x^\mu\mapsto x'^\mu=x^\mu-\xi^\mu(x)\,,
\label{skcdfjdkl}
\eeq
we find that the action transforms infinitesimally as
\beq
S\mapsto S+\int d^dx\partial_\mu(\xi^\mu{\cal L})\,.
\label{Var1}
\eeq
Alternatively, we may consider that the Lagrangian density ${\cal L}$ is a function of fields $\phi$ (not necessarily scalars) and their (first-order) derivatives ${\cal L}={\cal L}(\phi,\partial\phi)$. The fields $\phi$ vary under the  transformation \eqref{skcdfjdkl} infinitesimally as
\beq
\delta_\xi\phi=\phi'(x)-\phi(x)=\xi^\mu\partial_\mu\phi+
\hat\delta_\xi \phi\,,
\label{kckckddkd}
\eeq
where $\hat\delta_\xi$ includes any tensorial transformations of $\phi$.\footnote{Like as, a covariant or a contravariant tensor transform as
\begin{equation*}
\delta_\xi A_\mu=A_\mu'(x)-A_\mu(x)=\xi^\nu\partial_\nu A_\mu+
\partial_\mu\xi^\nu A_\nu\,,\quad
\delta_\xi B^\mu=B^{\mu'}(x)-B^\mu(x)
=\xi^\nu\partial_\nu B^\mu-
B^\nu\partial_\nu\xi^\mu\,,
\end{equation*}
as it can be easily proved by varying $A=A_\mu(x)dx^\mu$ and $B=B^\mu(x)\partial_\mu$.}
Under the latter the action transforms as the Lie derivative
\beq
S\mapsto S+\int d^dx\,\partial_\mu\left(\frac{\partial{\cal L}}{\partial(\partial_\mu \phi)}\delta_\xi\phi\right)\,,
\label{Var2}
\eeq
upon using the equations of motion. Matching the variations \eqref{Var1}, \eqref{Var2} and also using \eqref{kckckddkd} we find the generalised Noether current $S^\mu$
\beq
\partial_\mu S^\mu=0\,,\quad
S^\mu=\xi^\nu t^\mu{}_\nu-
\frac{\partial{\cal L}}{\partial(\partial_\mu \phi)}\hat\delta_\xi\phi\,,
\label{jnfsksdqw}
\eeq
where $t^\mu{}_\nu$ is the canonical conserved stress energy tensor given by
\beq
t^\mu{}_\nu=-\frac{\partial {\cal L}}{\partial(\partial_\mu\phi)}\partial_\nu\phi+\delta^\mu{}_\nu{\cal L}\,.
\eeq

For translations $\xi^\mu=a^\mu$ and a scalar field $\phi$ equation \eqref{jnfsksdqw} leads to the conservation of the canonical stress energy tensor
\beq
\partial_\mu t^\mu{}_\nu=0\,.
\eeq
In what follows we will work out three known cases where the additional term 
in \eqref{jnfsksdqw} contribute.

\subsection*{Scale}

Firstly, we consider the free scalar \eqref{action.free} for scale transformations 
\beq
x'^\mu=\lambda^{-1}x^\mu\,,\quad
\xi^\mu=\varepsilon x^\mu\,,
\quad
\phi'(x')=\det\left(\frac{\partial x'}{\partial x}\right)^{-\frac{\Delta_\phi}{d}}\phi(x)=\lambda^{-\Delta_\phi}\phi(x)\,,
\eeq
where $\lambda=1+\varepsilon$ and $\Delta_\phi=\frac{1}{2}(d-2)$. The additional term in \eqref{kckckddkd} can be easily found to be
\beq
\hat \delta_\xi\phi=-\varepsilon\Delta_\phi\phi
\eeq
and upon inserting into \eqref{jnfsksdqw} we obtain the on-shell conserved current
\beq
\partial_\mu S^\mu=0\,,\quad
S^\mu=x^\nu t^\mu{}_\nu +\frac12\Delta_\phi\partial^\mu\phi^2\,,
\eeq
where $t^{\mu\nu}$ was given in \eqref{emconfscalar} and the current precisely matches \eqref{skkn0} and \eqref{skkn1}.

\subsection*{Special conformal}

Secondly, we consider the action of special conformal transformations on the free scalar \eqref{action.free}
\beq
\label{sjdksl}
\begin{split}
&x'^\mu=\Omega^{-1}(x^\mu+b^\mu x^2)\,,\quad ds'^2=\eta_{\mu\nu}dx'^\mu dx'^\nu=\Omega^{-2}\eta_{\mu\nu}dx^\mu dx^\nu\,,\\
&\phi'(x')= \det\left(\frac{\partial x'}{\partial x}\right)^{-\frac{\Delta_\phi}{d}}\phi(x)=\Omega^{-\Delta_\phi}\phi(x)\,,\quad \Omega=1+2 b\cdot x+x^2b^2\,,
\end{split}
\eeq
where $\Delta_\phi=\frac{1}{2}(d-2)$.
The additional term in \eqref{kckckddkd} can be easily found to be
\beq
\hat\delta_\xi\phi=-2\Delta_\phi x_\nu b^\nu\phi\,,\quad
\xi^\mu=b^\mu x^2-2b^\nu x^\mu x_\nu\,.
\eeq
Upon using the latter into the  condition \eqref{jnfsksdqw}, we find that it is not on-shell conserved, satisfying
\beq
\partial_\mu S^\mu=-\Delta_\phi b^\mu \partial_\mu\phi^2\quad\text{where}\quad
S^\mu=\xi^\rho t^\mu{}_\rho-\Delta_\phi\partial^\mu\phi^2 b^\rho x_\rho \,.
\eeq
Yet this divergence can be readily written as a conserved quantity 
\beq
\partial_\mu(S^\mu+\Delta_\phi b^\mu\phi^2)=0\quad\Longrightarrow\quad
\partial_\mu\left(x^2t^\mu{}_\rho-2x^\nu t^\mu{}_\nu x_\rho+2(\partial_\nu Z^{\mu\nu} x_\rho-Z^\mu{}_\rho)\right)=0\,,
\eeq
where
\beq
Z^{\mu\nu}=-\frac12\Delta_\phi\eta^{\mu\nu}\phi^2\,,
\eeq
which is in agreement with \eqref{skkn20} and \eqref{current.special.conformal}.

The origin of the additional term can be understood upon transforming the free scalar \eqref{action.free} under the special conformal transformations
\eqref{sjdksl}, which is invariant but up to a boundary term~\cite{Wess:1960}
\beq
\begin{split}
S_\text{scalar}&=-\frac12\int d^dx\,\partial_\mu \phi\partial^\mu\phi\quad\Longrightarrow\quad\\ 
S'_\text{scalar}&=-\frac12\int d^dx\,\Omega^{2-d}\,\partial_\mu( \Omega^{\Delta_\phi}\phi)\partial^\mu( \Omega^{\Delta_\phi}\phi)\\ 
&=-\frac12\int d^d x\left\{(\partial\phi)^2+\Delta_\phi\partial_\mu\ln\Omega\,\partial^\mu\phi^2+\Delta_\phi^2(\partial\ln\Omega)^2\phi^2\right\}\\
&=-\frac12\int d^d x\left\{(\partial\phi)^2+\Delta_\phi\partial_\mu\left(\partial^\mu\ln\Omega\,\phi^2\right)\right\}\,,\\
&=S_\text{scalar}-\frac{\Delta_\phi}{2}\int d^d x\,\partial_\mu\left(\phi^2\partial^\mu\ln\Omega\right)\,,
\end{split}
\eeq
where in the second to last equation we have used the following identity that is valid for the special conformal transformations \eqref{sjdksl}
\beq
\Delta_\phi(\partial\ln\Omega)^2=\partial^2\ln\Omega\,.
\eeq

\subsection*{Lorentz}

Thirdly, we consider the Maxwell field \eqref{Maxwell.action} which is invariant under Lorentz transformations 
\beq
\xi^\nu=\omega^\nu{}_\rho x^\rho\,,\quad 
\hat\delta_\xi A_\mu=-\omega_\mu{}^\nu{} A_\nu\,,
\eeq
where $\omega_{\mu\nu}$ is an antisymmetric tensor. Using the above and \eqref{canonicalEM} into \eqref{jnfsksdqw} we find
\beq
\partial_\mu J^{\mu\nu\rho}=0\,,\quad
J^{\mu\nu\rho}=x^\rho t^{\mu\nu}-x^\nu t^{\mu\rho}
+F^{\mu\rho}A^\nu-
F^{\mu\nu}A^\rho\,,
\eeq
that is in agreement with \eqref{fksksaal} and \eqref{qgeuwuw}.

\end{document}